\documentclass[a4paper,11pt]{article}
\pdfoutput=1 

\usepackage[no-natbib-sort]{jheppub} 

\usepackage[T1]{fontenc} 

\usepackage{todonotes}

\natbibsortfalse

\title{\boldmath Abelian F-theory Models with Charge-3 and Charge-4 Matter}

\author[a]{Nikhil Raghuram}

\affiliation[a]{Center for Theoretical Physics\\Department of Physics\\Massachusetts Institute of Technology\\77 Massachusetts Avenue\\Cambridge, MA 02139, USA}

\emailAdd{{\tt nikhilr} {\rm at} {\tt mit.edu}}

\preprint{MIT-CTP-4961}

\newcommand{\kmopr}[0]{KMOPR } 
\newcommand{\quotring}[1]{R/\langle#1\rangle}
\newcommand{\normring}[1]{\widetilde{R/\langle#1\rangle}}
\newcommand{\seccoord}[1]{\hat{#1}} 
\newcommand{\secx}[0]{\seccoord{x}} 
\newcommand{\secy}[0]{\seccoord{y}} 
\newcommand{\secz}[0]{\seccoord{z}} 
\newcommand{\ratsec}[1]{\hat{#1}}
\newcommand{\zerosec}[0]{\ratsec{o}}
\newcommand{\sechomol}[1]{\mathcal{#1}}
\newcommand{\zerohomol}[0]{\sechomol{Z}}
\newcommand{\gu}[0]{\mathrm{U}}
\newcommand{\gsu}[0]{\mathrm{SU}}
\newcommand{\gsp}[0]{\mathrm{Sp}}
\newcommand{\eplus}[0]{[+]}

\DeclareMathOperator{\Res}{Res}

\newcommand{\falsedisc}[0]{\delta}
\newcommand{\ba}[0]{b_{(2)}}
\newcommand{\bb}[0]{b_{(1)}}
\newcommand{\bc}[0]{b_{(0)}}
\newcommand{\betaa}[0]{\beta_{(2)}}
\newcommand{\betac}[0]{\beta_{(0)}}
\newcommand{\dbla}[0]{\eta_a}
\newcommand{\dblb}[0]{\eta_b}
\newcommand{\zL}[0]{\tilde{B}}
\newcommand{\cthree}[0]{t}
\newcommand{\thetaa}[0]{\theta_{(1)}}
\newcommand{\thetab}[0]{\theta_{(0)}}
\newcommand{\taua}[0]{\tau_{(2)}}
\newcommand{\taub}[0]{\tau_{(1)}}
\newcommand{\tauc}[0]{\tau_{(0)}}
\newcommand{\ta}[0]{t_{(3)}}
\newcommand{\tb}[0]{t_{(2)}}
\newcommand{\tc}[0]{t_{(1)}}

\newcommand{\td}[0]{t_{(0)}}
\newcommand{\PsiL}[0]{\tilde{\Xi}}
\newcommand{\phibar}[0]{\overline{\phi}}
\newcommand{\ordertworem}[0]{\falsedisc_{2,\text{rem}}}
\newcommand{\ctwo}[0]{\phi}
\newcommand{\phia}[0]{\phi_{(1)}}
\newcommand{\phib}[0]{\phi_{(0)}}
\newcommand{\orderthreerem}[0]{\falsedisc_{3,\text{rem}}}

\newcommand{\tausq}[0]{\tau_{\text{sq}}}
\newcommand{\taucu}[0]{\tau_{\text{cu}}}
\newcommand{\orderfourrem}[0]{\falsedisc_{4,\text{rem}}}

\newcommand{\fonesec}[0]{f_1}

\newcommand{\ha}[0]{h_{(2)}}
\newcommand{\hb}[0]{h_{(1)}}
\newcommand{\hc}[0]{h_{(0)}}
\newcommand{\lambdaa}[0]{\lambda_{(1)}}
\newcommand{\lambdab}[0]{\lambda_{(0)}}
\newcommand{\hcombo}[0]{h}
\newcommand{\mpa}[0]{a}

\newcommand{\mthree}[0]{m_3}
\newcommand{\mtwo}[0]{m_2}
\newcommand{\mone}[0]{m_1}

\newcommand{\qthreea}[0]{\alpha}
\newcommand{\qthreeb}[0]{\beta}
\newcommand{\cthreeqfour}[0]{t}
\newcommand{\rthree}[0]{r_{3}}
\newcommand{\rtwo}[0]{r_{2}}
\newcommand{\rone}[0]{r_{1}}

\newcommand{\ftwelveh}{\hat{f}_{12}}
\newcommand{\secgen}[0]{\ratsec{s}}
\abstract{This paper analyzes $\gu(1)$ F-theory models admitting matter with charges $q=3$ and $4$. First, we systematically derive a $q=3$ construction that generalizes the previous $q=3$ examples. We argue that $\gu(1)$ symmetries can be tuned through a procedure reminiscent of the $\gsu(N)$ and $\gsp(N)$ tuning process. For models with $q=3$ matter, the components of the generating section vanish to orders higher than 1 at the charge-3 matter loci. As a result, the Weierstrass models can contain non-UFD structure and thereby deviate from the standard Morrison-Park form. Techniques used to tune $\gsu(N)$ models on singular divisors allow us to determine the non-UFD structures and derive the $q=3$ tuning from scratch. We also obtain a class of a $q=4$ models by deforming a prior $\gu(1)\times\gu(1)$ construction. To the author's knowledge, this is the first published F-theory example with charge-4 matter. Finally, we discuss some conjectures regarding models with charges larger than 4.}

\begin{document} 
\maketitle
\flushbottom
\section{Introduction}
A key objective of the F-theory program is determining which charged matter representations can arise in F-theory models, a task with important implications for the landscape and swampland.  Clearly, we cannot characterize the full landscape of F-theory models without knowing all of the representations that can be realized in F-theory. At the same time, one may find that certain representations cannot be obtained in F-theory, even when the corresponding matter spectra satisfy the known low-energy conditions. This scenario would inspire a variety of questions, such as whether these representations could be attained through other string constructions or whether some previously unknown low-energy condition could explain the absence of these representations. And from a more mathematical perspective, exploring F-theory compactifications with different representations can tell us about the scope of Calabi-Yau geometries. Because of these ramifications, there has been much interest in developing techniques for building F-theory models with various matter spectra. For non-abelian groups, this line of inquiry has led to F-theory constructions admitting a wide range of representations \cite{katz-vafa,matter-singularities,esole-yau, ckpt, KleversTaylor, kmrt}. Abelian constructions and their matter spectra have been a focus of the F-theory literature as well, both in purely abelian situations and in contexts with additional non-abelian groups \cite{Grimm-Weigand-protondecay,Marsano-Saulina-Schafer-Nameki,Dolan-Marsano-Saulina-Schafer-Nameki,ParkTaylor,park-intersection,morrison-park,CveticGrimmKlevers,MayrhoferPaltiWeigand,cvetic-klevers-piragua,BraunGrimmKeitelII,CveticGrassiKleversPiragua,BorchmannMayrhoferPaltiWeigand,cvetic-u1cubed,AntoniadisLeontaris,KuntzlerSchaferNameki,kmpopr,EsoleKangYau,Lawrie-Sacco,f-theory-rational,ckpt,KleversTaylor,morrison-park-2016,MayorgaPena-Valandro,Cvetic-Lin,Buchmuller-Dierigl-Oehlmann-Ruehle}. In fact, classifying the possible charges of abelian F-theory models has an additional phenomenological importance given the role extra $\gu(1)$'s play in F-theory GUT model building \cite{Grimm-Weigand-protondecay,Dolan-Marsano-Schafer-Nameki,Braun-Grimm-Keitel,Krippendorf-SchaferNameki-Wong}. Nevertheless, the issue of how to construct an F-theory model with a desired abelian charge spectrum remains challenging, even for models with only a $\gu(1)$ gauge group. In particular, there are open questions regarding the construction of models with charges $q>2$ (in appropriately quantized units). The goal of this work is to provide new insights into F-theory models admitting $q=3$ and $q=4$ matter, with the hope that these ideas can inform our understanding of models with arbitrary charges.

The reason for the more challenging nature of abelian F-theory models lies in the different manifestations of non-abelian and abelian symmetries. F-theory models in $12-2d$ dimensions are constructed using a Calabi-Yau $d$-fold that is an elliptic fibration over a base $B$. Non-abelian gauge symmetries occur when the fiber becomes singular along a codimension one locus in $B$, while charged matter often occurs at codimension two loci with singular fibers. The codimension one singularity types and their corresponding non-abelian gauge algebras have already been classified \cite{morrison-vafa-I,morrison-vafa-II,bershadsky-etal,tate-f-theory}, and in many cases, one can relate the codimension two singularity types to different charged matter representations \cite{sadov,katz-vafa,matter-singularities}. These dictionaries provide a strategy for constructing an F-theory model admitting a particular gauge group and charged matter spectrum. One first reads off the singularity types and loci that produce the desired gauge data. Then, one determines the algebraic conditions that make the elliptic fibration support the appropriate singularities. This process, known as tuning, has been used to systematically construct a variety of non-abelian gauge groups and charged matter \cite{matter-singularities,matter-in-transition,kmrt}.

In contrast, abelian gauge groups are not associated with elliptic curve singularities along codimension one loci. They instead arise when there are additional rational sections of the elliptic fibration, such that the elliptic fibration has a non-trivial Mordell-Weil group \cite{morrison-vafa-II,park-intersection,morrison-park}. Thus, the usual procedures for obtaining non-abelian groups do not carry over to abelian groups in an immediately obvious way, making the construction of F-theory models with abelian gauge symmetries more difficult. Take, for example, the question of how to construct an F-theory model with a single $\gu(1)$ gauge group and no additional non-abelian groups. There is a well known $\gu(1)$ construction, the Morrison-Park model \cite{morrison-park}, but it admits only $q=1$ and $q=2$ matter. \cite{kmpopr} presented a construction supporting $q=3$ matter, which was found within a set of toric models. However, this construction was found somewhat by chance, raising the question of whether it could be systematically derived from scratch. That is, instead of looking within a set of models, could someone start with the goal of finding a $q=3$ model and follow a series of steps to obtain this construction? The Weierstrass model also has a structure quite different from the Morrison-Park form, posing the related question of whether we can understand how and why the structures differ. While there has been some discussion of F-theory models with $q=4$ matter \cite{oehlmann-private}, there is, to the author's knowledge, no published $\gu(1)$ model with charges $q\geq 4$. This makes an understanding of $q=3$ models all the more important, as the features that distinguish the $q=3$ construction from the Morrison-Park form would likely play a role in $q\geq4$ models as well.

This work presents a systematic method for tuning a $q=3$ construction and presents a class of models admitting $q=4$ matter. A central theme is that the presence of $q\geq3$ matter is tied to the order of vanishing of the section components. As is well known from \cite{morrison-park}, $q=2$ matter occurs when the components of the section vanish on some codimension two locus; in Weierstrass form, the $\secz$, $\secx$, and $\secy$ components vanish to orders 1, 2, and 3. In the models discussed here, the section components vanish to higher orders at the $q\geq 3$ loci, directly affecting the structure of the Weierstrass model. For instance, the $\secz$ component of the $q=3$ construction vanishes to order 2 on the $q=3$ locus, reminiscent of a divisor with double point singularities. As discussed in Section \ref{sec:charge-3-models}, one can build abelian F-theory models through a process similar to the $\gsu(N)$ and $\gsp(N)$ tuning procedure. Instead of making the discriminant proportional to a divisor supporting a non-abelian symmetry, we tune quantities to be proportional to the $\secz$ component of the section. When $\secz$ vanishes to orders larger than 1, the tuning process allows for structures associated with rings that are not unique factorization domains (UFDs); these structures can be derived using the normalized intrinsic ring technique of \cite{kmrt}. Following the procedure leads to a generalization of the previous $q=3$ construction in \cite{kmpopr}, with a direct link between the specific structures in the $q=3$ Weierstrass model and the singular nature of $\secz$. We also obtain a $q=4$ F-theory construction by deforming a previous $\gu(1)\times\gu(1)$ construction from \cite{ckpt}. To the author's knowledge, this is the first published F-theory example admitting $q=4$ matter. While we do not derive this construction using the normalized intrinsic ring, the section components of the $q=4$ construction vanish to higher orders as well, and the Weierstrass model contains structures suggestive of non-UFD behavior.

The rest of this paper is organized as follows. Section \ref{sec:f-theory-overview} reviews some aspects of abelian groups in F-theory that are important for the discussion. Section \ref{sec:charge-3-models} describes how abelian symmetries can be tuned and uses the process to systematically derive a $q=3$ construction. In Section \ref{sec:charge-4-models}, we construct and analyze a construction admitting $q=4$ matter. Section \ref{sec:higher-charges} includes some comments about $q>4$ models, while Section \ref{sec:conclusions} summarizes the findings and mentions some directions for future work. There are accompanying Mathematica files containing expressions for the constructions derived here; details about these Mathematica files are given in Appendix \ref{app:mathematicafiles}.

\section{Overview of abelian gauge groups in F-theory}
\label{sec:f-theory-overview}
In this section, we review those aspects of F-theory that are necessary for the rest of the discussion. We will not be too detailed here, instead referring to the mentioned references for further details. More general reviews of F-theory can be found in \cite{denef-review,weigand-review,taylor-review}.

F-theory can be described from either a Type IIB perspective or an M-theory perspective. In the Type IIB view, an F-theory model can be thought of as a Type IIB compactification in which the presence of 7-branes causes the axiodilaton to vary over the compactification space. The axiodilaton is represented as the complex structure of an elliptic curve, and the F-theory compactification involves an elliptic fibration $X$ over a compactification base $B$. In this paper, we will assume that the base $B$ is smooth. Mathematically, the elliptic fibration can be described using the global Weierstrass equation
\begin{equation}
y^2 = x^3 + f x z^4 + g z^6. \label{eq:globWeierstrassform}
\end{equation}
$[x:y:z]$ refer to the coordinates of a $\mathbb{P}^{2,3,1}$ projective space in which the elliptic curve is embedded, and $f$ and $g$ are sections of line bundles over $B$. To guarantee a consistent compactification that preserves some supersymmetry, we demand that the total elliptic fibration $X$ is a Calabi-Yau manifold by imposing the Kodaira constraint: $f$ and $g$ must respectively be sections of $\mathcal{O}(-4K_B)$ and $\mathcal{O}(-6K_B)$, where $K_B$ is the canonical class of the base $B$. The Weierstrass equation is often written in a chart where $z\neq 0$, in which case the $x,y,z$ coordinates can be rescaled so that $z=1$. This procedure leads to the local Weierstrass form
\begin{equation}
y^2 = x^3 + f x + g \label{eq:locWeierstrassform}
\end{equation}
commonly seen in the F-theory literature. Note that the elliptic fiber is allowed to be singular along loci in the base. Codimension one loci with singular fibers are associated with non-abelian gauge groups, while codimension two loci with singular fibers are associated with charged matter.

F-theory can also be understood via its duality with M-theory. To illustrate the idea, let us first consider M-theory on $T^2$. Shrinking one of the cycles in the $T^2$ leads to Type IIA compactified on $S^1$, which is dual to Type IIB on $S^1$. The radii of the circles in the dual Type II theories are inverses of each other, and if we shrink the Type IIA circle, the circle dimension on the Type IIB side decompactifies. Similarly, we can consider M-theory on a smooth, elliptically fibered CY $d$-fold. Roughly, applying the above shrinking procedure fiberwise gives a Type IIB theory on the base $B$ with a varying axiodilaton $\tau$. This Type IIB model can then be thought of as an F-theory model on an elliptically fibered CY $d$-fold. Of course, the full duality involves several subtleties not captured in the discussion above, particularly with regards to singularities and the details of the shrinking procedure. While these issues are not too crucial for the discussion here, readers interested in further details can consult, for instance, \cite{grimm-4dmfdual,bonetti-grimm}.

\subsection{Elliptic curve group law}
\label{subsec:additionlaw}

The ultimate goal of this section is to describe rational sections of elliptic fibrations and their relation to the abelian sector of F-theory models. However, it is helpful to first describe the addition law on elliptic curves, as it plays an important role in the discussion. This subsection is largely based on \cite{silverman}, to which we refer for further details.

The points of an elliptic curve form an abelian group under an addition operation that we denote $\eplus$. To describe the addition law, we first identify a particular point $Z$ as the identity of the group. Given two points $P$ and $Q$, we find $P\eplus Q$ by first forming a line that passes through both $P$ and $Q$; if $P$ and $Q$ are the same point, we instead form the tangent line to the elliptic curve at $P$. This line intersects the elliptic curve at a third point $R$. We then form the line that passes through $R$ and the identity point $Z$ (or if $Z=R$, the tangent line to the elliptic curve at $Z$). This second line again intersects the elliptic curve at a third point, which is taken to be $P\eplus Q$. One can show that the addition law satisfies all of the axioms for an abelian group. In particular, the inverse of a point $P$, which is denoted as $-P$, is found through the following procedure. First, we form the tangent line to the elliptic curve at $Z$, which intersects the elliptic curve at a point $S$. Then, $-P$ is the third intersection point of the line passing through $S$ and $P$. 

It is useful to have explicit expressions for the addition law when the elliptic curve is written in the global Weierstrass form \eqref{eq:globWeierstrassform}. The identity element $Z$ is typically chosen to be the point $[x:y:z]=[1:1:0]$. Note that, in Weierstrass form, $Z$ is a flex point\footnote{While $Z$ is a flex point in Weierstrass form, the identity element may not be a flex point when an elliptic curve is written in other forms. This subtlety is particularly relevant for the $\mathbb{P}^2$ form of the $q=4$ elliptic fibration in \S\ref{sec:charge-4-models}.}, as the tangent line at $Z$ intersects the elliptic curve at this point with multiplicity 3; in other words, the tangent line at $Z$ does not intersect the elliptic curve at any point other than $Z$. Given two points $P=[x_P:y_P:z_P]$ and $Q=[x_Q:y_Q:z_Q]$, $P\eplus Q$ has coordinates\footnote{If desired, one could use the Weierstrass equation to eliminate $f$ and $g$ and rewrite \eqref{eq:weieraddlawx} through \eqref{eq:weieraddlawz} entirely in terms of the $P$ and $Q$ coordinates. Additionally, the elliptic curve addition formula is typically written in a chart where $z=1$. After setting $z_P$ and $z_Q$ to 1 in the expressions and eliminating $f$ and $g$, one recovers the standard form given in, for example, Appendix A of \cite{morrison-park}.}
\begin{align}
x =& x_P z_P^2\left(x_Q^2 + f z_Q^4\right) + x_Q z_Q^2\left(x_P^2 + f z_P^4\right)-2 z_P z_Q \left(y_P y_Q - g z_P^3 z_Q^3\right) \label{eq:weieraddlawx}\\ 
y =& -y_P^2 y_Q z_Q^3 -3 x_Q x_P^2 y_Q z_Q z_P^2+3 x_P x_Q^2 y_P z_P z_Q^2+y_Q^2 y_P z_P^3-3 g z_P^3 z_Q^3\left(y_Q z_P^3 - y_P z_Q^3\right) \notag\\
&- f z_P z_Q\left(x_Q y_Q z_P^5 + 2 x_P y_Q z_P^3 z_Q^2 - 2 x_Q y_P z_Q^3 z_P^2 - x_P y_P z_Q^5\right)\label{eq:weieraddlawy}\\
z=& x_Q z_P^2 - x_P z_Q^2.\label{eq:weieraddlawz}
\end{align}
Meanwhile, the point $P\eplus P = 2P$ has the coordinates
\begin{align}
x =& \left(3 x_P^2 + f z_P^4\right)^2 -8 x_P y_P^2 \\
y=& -\left(3 x_P^2 + f z_P^4\right)^3 + 12 x_P y_P^2 \left(3 x_P^2 + f z_P^4\right) - 8 y_P^4\\
z =& 2 y_P z_P.
\end{align}
Note that the $2P$ expressions do not follow directly from plugging $z_Q=z_P, x_Q= x_P, y_Q=y_P$ into \eqref{eq:weieraddlawx} through \eqref{eq:weieraddlawz}, as all of the section components in \eqref{eq:weieraddlawx}-\eqref{eq:weieraddlawz} vanish with this substitution. For a point $P=[x_P, y_P, z_P]$, the inverse $-P$ is simply $[x_P:-y_P:z_P]$.

\subsection{Rational sections, the abelian sector, and the Mordell-Weil group}
\label{subsec:mordellweil}

Unlike the non-abelian sector, the abelian sector of the gauge group is not associated with codimension one loci in the base with elliptic curve singularities. Instead, the abelian sector is associated with rational sections of the elliptic fibration. 

For our purposes, an F-theory construction will always have at least one rational section, the zero section $\zerosec$.\footnote{See \cite{BraunMorrison,mt-sections,Anderson-multisection,cvetic-z3} for discussions of situations without a zero section.} If the model is written in the global Weierstrass form of Equation \eqref{eq:globWeierstrassform},  the zero section is
\begin{equation}
\zerosec: [\secx:\secy:\secz] = [1:1:0].
\end{equation}
But an elliptic fibration may have additional rational sections. In fact, these rational sections form a group, known as the Mordell-Weil group, under the addition operation described in \S\ref{subsec:additionlaw}, with $\zerosec$ serving as the identity \cite{Wazir}. According to the Mordell-Weil theorem \cite{Neron-Lang}, the group is finitely generated and takes the form
\begin{equation}
\mathbb{Z}^r \oplus \mathcal{G}.
\end{equation}
$\mathcal{G}$ is the torsion subgroup, with every element of $\mathcal{G}$ having finite order; the torsion group will not be important for the purposes of this paper. $r$ meanwhile is called the Mordell-Weil rank.

If an elliptic fibration has Mordell-Weil rank $r$, the abelian sector of the corresponding F-theory model includes a $\gu(1)^r$ gauge algebra \cite{morrison-vafa-II,park-intersection,morrison-park}. The justification for this statement is most easily seen in the dual M-theory picture, as discussed in \cite{park-intersection}. For concreteness, let us restrict ourselves to 6D F-theory models, although similar arguments apply in 4D. Additionally, we assume there are no codimension one singularities apart from the standard $I_1$ singularity, as we are not interested in situations with non-abelian symmetry. Consider M-theory compactified on a resolved elliptically fibered Calabi-Yau threefold $\tilde{X}$. M-theory on $\tilde{X}$ is a 5D model that, in the F-theory limit, leads to a 6D $N=1$ F-theory model. According to Poincar\'{e} duality, there is a harmonic two-form $\omega$ for every four-cycle $\Sigma$ in $\tilde{X}$. The two-forms serve as zero-modes for the M-theory three-form  $C_3$, and we can expand $C_3$ using a basis of two-forms. In other words, we write $C_3$ as a sum of terms of the form $A\wedge \omega$; the one-forms $A$ represent vectors in the 5D theory. Thus, to find the vectors of the 6D F-theory model, we consider a basis of four-cycle homology classes of $\tilde{X}$, find the corresponding 5D vectors $A$, and track the sources of these 5D vectors in the 6D F-theory model.

When there are no codimension one singularities (apart from $I_1$ singularities), there are three types\footnote{When there are codimension one singularities, there is a fourth type of four-cycle homology class that corresponds to the Cartan gauge bosons of a non-abelian gauge group in the F-theory model. Since we are not interested in the possibility of additional non-abelian gauge groups here, we ignore this fourth type of four-cycle. See \cite{park-intersection} for further details.} of four-cycle homology classes that are of interest: the homology class $\zerohomol$ associated with the zero section, the homology classes $\sechomol{S}_1$ through $\sechomol{S}_r$ associated with the $r$ generators of the Mordell-Weil group, and the homology classes $B_\alpha$ that come from fibering the elliptic curve over two-cycles in the base. 5D vectors associated with $\zerohomol$ and $B_\alpha$ do not correspond to gauge bosons in the 6D F-theory model. Instead, they arise from the KK reduction of either the metric or tensors in the 6D F-theory model. But 5D vectors associated to $\sechomol{S}_1$ through $\sechomol{S}_r$ come from vector multiplets in the 6D model. These are the gauge bosons for the $\gu(1)^r$ gauge group.

However, the 5D vectors do not directly correspond to the $\sechomol{S}_i$ but are rather associated with combinations of $\sechomol{S}_i$ with $\zerohomol$ and the $B_\alpha$. At least informally, we must isolate the part of the $\sechomol{S}_i$ that is orthogonal to the other four-cycles. This is done using the Tate-Shioda map $\sigma$, which is a homomorphism from the Mordell-Weil group to the homology group of four-cycles. For a situation with no codimension one singularities, the Tate-Shioda map is given by \cite{morrison-park}
\begin{equation}
\sigma(\ratsec{s}) = \sechomol{S} - \zerohomol - \left(\sechomol{S}\cdot\zerohomol\cdot B^{\alpha} - K_B^{\alpha}\right)B_{\alpha},
\end{equation}
where $K_B^{\alpha}$ are the coordinates of the canonical class of the base written in the basis $B_{\alpha}$. Thus, the $\gu(1)$ gauge bosons are actually associated with the homology class $\sigma(\ratsec{s}_i)$, and the Tate-Shioda map plays an important role in physical expressions.

An important property of a rational section $\ratsec{s}$, particularly for anomalies, is its height $h(\ratsec{s})$. The height is a divisor in the base given by \cite{morrison-park}
\begin{equation}
h(\ratsec{s}) = -\pi\left(\sigma(\ratsec{s})\cdot\sigma(\ratsec{s})\right),
\end{equation}
where $\pi$ is a projection onto the base. For a 6D F-theory model with no codimension one singularities apart from $I_1$ singularities, the height can be expressed in a simpler form \cite{morrison-park,morrison-park-2016}:
\begin{equation}
h(\ratsec{s}) = 2\left(-K_B + \pi(\sechomol{S}\cdot\zerohomol)\right),
\end{equation}
where $\sechomol{S}$ is the homology class of the section $\ratsec{s}$. This expression can often be simplified further. Suppose that, in global Weierstrass form, the section has coordinates $[\secx:\secy:\secz]$. Additionally, assume that the coordinates have been scaled so that they are all holomorphic and that there are no common factors between $\secx$, $\secy$ and $\secz$ that could be removed by rescalings. We can consider a curve $\secz=0$ in the base, and we denote the homology class of this curve $[\secz]$. $\ratsec{s}$ coincides with the zero section at loci in the base where $\secz=0$, so the height is given by \cite{morrison-park,morrison-park-2016}
\begin{equation}
h(\ratsec{s}) = 2\left(-K_B + [\secz]\right). \label{eq:heightsec}
\end{equation}
Since the height is written entirely in terms of homology classes of the base, this expression is useful for calculations, particularly those related to anomaly cancellation. Note that if there are multiple generators, one may be interested in a height matrix, which includes entries such as $ -\pi\left(\sigma(\ratsec{s}_i)\cdot\sigma(\ratsec{s}_j)\right)$ for distinct generators $\ratsec{s}_i$ and $\ratsec{s}_i$. Here, we are primarily interested in situations with a rank-one Mordell Weil group, so this generalized form will not be too important. 

\subsection{Charged matter}

Even though the abelian gauge symmetry is not associated with codimension one singularities, charged matter still occurs at codimension two loci with singular fibers, as discussed in \cite{park-intersection}. Again, we restrict ourselves to a model with an abelian gauge group but no additional non-abelian gauge groups. The model has various codimension two loci with $I_2$ singularities. After these singularities are resolved, the fibers at these codimension two loci consist of two $\mathbb{P}^1$s which intersect each other at two points. One of the components, the one containing the zero section, can be thought of as the main elliptic curve, with the other component being the extra $\mathbb{P}^1$ introduced to resolve the singularity. In the M-theory picture, charged matter arises from M2 and anti-M2 branes wrapping this extra component.

To calculate the charge of this matter, we must examine the M2 brane world-volume action. The action contains a term of the form $\int C_3$, where the integral is over the M2 brane world-volume. For the situation at hand, the M2 brane wraps a component $c$ of the singular fiber. $C_3$ meanwhile has an expansion involving terms of the form $A\wedge \omega$, where $\omega$ is a harmonic two-form of the resolved CY manifold $\tilde{X}$. Integrating over the $c$ component leads to a term in the action of the form  $\int A$ over a world-line, thereby giving the action for charged matter. The charge comes from integrating the two-form $\omega$ associated with the $\gu(1)$ gauge boson $A$. However, for a CY $n$-fold, each $\omega$ is dual to a $(2n-2)$-cycle $\Sigma$, and for any two-cycle $c$,
\begin{equation}
\int_c \omega = c \cdot \Sigma.
\end{equation}
The gauge boson $A$ for a generator $\ratsec{s}$ in the Mordell-Weil group is associated with $\sigma(\ratsec{s})$. Therefore, the charges supported at an $I_2$ locus are given by
\begin{equation}
q = \pm \sigma(\hat{s})\cdot c.
\end{equation}
The sign corresponds to whether $c$ is wrapped by an M2 brane or an anti-M2 brane. In situations without additional non-abelian symmetries, the charge formula reduces to \cite{park-intersection,morrison-park}
\begin{equation}
q=\left(\sechomol{S}-\zerohomol\right)\cdot c.
\end{equation}

For a generating section $\secgen = [\secx:\secy:\secz]$, charged matter occurs at \cite{morrison-park,cvetic-klevers-piragua}
\begin{equation}
\secy = 3\secx^2 + f\secz^4 =0. \label{eq:generalmatterlocus}
\end{equation}
Clearly, the above condition is satisfied if all of the components of the section vanish at some codimension two locus. Not only is the elliptic fiber singular when this happens, but the section itself is ill-defined. Analyzing such situations requires that we resolve the section, a process described in \cite{morrison-park}. Afterwards, the section appears to ``wrap'' one of the $\mathbb{P}^1$'s of the $I_2$ fiber. Rational sections typically behave this way at loci supporting $q\geq 2$ matter. At $q=2$ loci, the $\secz$, $\secx$, and $\secy$ components (in Weierstrass form) vanish to orders 1,2, and 3. As described later, the components vanish to higher orders at loci supporting $q\geq 3$ matter. For instance, $\secz$ vanishes to order 2 for $q=3$ loci and order 4 for $q=4$ loci. This higher order of vanishing likely affects the way the section wraps components, but we will not significantly investigate resolutions of the $q=3$ and $q=4$ models here. However, it would be interesting to better understand the wrapping behavior in models with $q\geq 3$ matter in future work. 

\subsection{Anomaly cancellation}

Any F-theory construction should satisfy the low-energy anomaly cancellation conditions from supergravity. Since 6D is the largest dimension in which supergravity theories can admit charged matter, the 6D anomaly cancellation conditions will be particularly important here as a consistency check on the models. In 6D supergravity models, anomalies are typically canceled through the Green-Schwarz mechanism. However, not all models are anomaly free; in order for anomalies to cancel, the massless spectrum must obey particular conditions. While the anomaly cancellation conditions come from low-energy considerations, they do have a geometric interpretation in F-theory \cite{park-intersection}, and the conditions can be written in terms of parameters describing the F-theory compactification.

The general anomaly cancellation conditions for models with abelian gauge groups are given in \cite{Erler,ParkTaylor,park-intersection}. Here, we restrict our attention to the case of a single $\gu(1)$ gauge group with no additional gauge symmetries. In the F-theory model, the Mordell-Weil group is generated by a single section, which we refer to as $\ratsec{s}$. Suppose the model has a base $B$ with canonical class $K_B$. Then, the gauge and mixed gravitational-gauge anomaly conditions are
\begin{align}
-K_B \cdot h(\ratsec{s}) &= \frac{1}{6}\sum_{I}q_I^2 & h(\ratsec{s})\cdot h(\ratsec{s}) &=\frac{1}{3}\sum_{I}q_I^4. \label{eq:anom-cond}
\end{align}
The index $I$ runs over the hypermultiplets, with $q_I$ denoting the charge of the $I$th hypermultiplet. $h(\ratsec{s})$ meanwhile is the height of the section $\ratsec{s}$, as described in \ref{subsec:mordellweil}. There are also the pure gravitational anomaly conditions
\begin{align}
H - V + 29 T &= 273 & K_B \cdot K_B &= 9-T,
\end{align}
where $H$, $V$, and $T$ denote the total number of hypermultiplets, vector multiplets, and tensor multiplets, respectively. Again, the anomaly conditions can be viewed as fully low-energy supergravity constraints, even though they are phrased here in terms of F-theory parameters.

The anomaly conditions can be used to derive two relations that are particularly useful for $q\geq 3$ models. The first is the tallness constraint \cite{morrison-park-2016}
\begin{equation}
\frac{h(S)\cdot h(S)}{-2 K_B \cdot h(S)} \leq \max_{I} q_{I}^2.\label{eq:tallnessconstr}
\end{equation}
This constraint suggests that a section with large enough $h(\ratsec{s})$ is forced to have some higher charge matter. But the anomaly equations in \eqref{eq:anom-cond} also imply that\footnote{While this work was being completed, the author became aware of the upcoming work \cite{MonnierMoorePark}, which independently derives \eqref{eq:abeliangenusinit} as part of a broader analysis of 6D supergravity constraints. It features a more detailed analysis of this relation along with analogues for situations with multiple $\gu(1)$ factors.}
\begin{equation}
h(\ratsec{s})\cdot\left(h(\ratsec{s})+2 K_B\right) = \frac{1}{3}\sum_{I}q_I^2\left(q_I^2-1\right).\label{eq:abeliangenusinit}
\end{equation}
Specializing to situations where \eqref{eq:heightsec} applies, this relation can be rewritten as
\begin{equation}
[\secz]\cdot\left(-K_B + [\secz]\right) = \frac{1}{12}\sum_{I}q_I^2\left(q_I^2-1\right) \label{eq:abeliangenus}
\end{equation}
Note that $q^2(q^2-1)/12$ is 0 for $q=0,1$ and is a positive integer for $q\geq 2$. Anomalies therefore directly determine the number of $q=2$ hypermultiplets given $h(\ratsec{s})$, $K_B$, and the number of $q\geq 3$ multiplets; importantly, the $q=2$ multiplicity can be determined without any information about the $q=1$ hypermultiplets. As discussed in \S\ref{subsec:q3matter} and \S\ref{subsec:q4matter}, this anomaly relation seems to have a direct F-theory realization: it describes the loci where the three components of the section vanish, leaving the section ill-defined. Moreover, every term in the sum on the right-hand side is non-negative, allowing us to conclude that
\begin{equation}
h(\ratsec{s})\cdot\left(h(\ratsec{s})+2 K_B\right) \geq \max_{I}\frac{1}{3}q_I^2\left(q_I^2-1\right).
\end{equation}
This bound in some sense has the opposite effect as the tallness constraint: if we wish to obtain a model admitting a certain charge $q$, we must have a sufficiently large $h(\ratsec{s})$. The relation resembles the genus condition \cite{KumarParkTaylor} for $\gsu(2)$ F-theory models, although we leave an in-depth exploration of any connection to future work. 

\section{Charge-3 models}
\label{sec:charge-3-models}

While there is a previous F-theory construction admitting $q=3$ matter \cite{kmpopr}, there are still open questions regarding its intricate structure. On the one hand, the construction in \cite{kmpopr}, which we henceforth refer to as the \kmopr model, was not purposefully constructed with the goal of realizing $q=3$ matter. Instead, it was found somewhat by chance in a class of toric constructions. But if we wish to understand ways of obtaining $q>3$ models, it behooves us to determine whether we can construct $q=3$ models from scratch. That is, rather than searching through a set of constructions with the hope of finding a $q=3$ model, could we use general principles and mathematical conditions to directly construct a $q=3$ model? Moreover, \cite{KleversTaylor} argued that the structure of the \kmopr model differs from that of the well-known Morrison-Park construction \cite{morrison-park}. In \cite{morrison-park-2016}, it was shown that the \kmopr Weierstrass model is birationally equivalent to one in Morrison-Park form, although the Morrison-Park form Weierstrass model does not satisfy the Calabi-Yau condition. Nevertheless, the analysis in \cite{morrison-park-2016} depended on unexpected cancellations between expressions in the \kmopr model. \cite{KleversTaylor, morrison-park-2016} hinted that the cancellations could be explained using rings that are not unique factorization domains (UFDs), but they did not describe how to understand or derive the construction's specific structures. 

This section describes a method for systematically deriving a $q=3$ construction. One can construct a Weierstrass model with non-trivial Mordell-Weil rank through a process similar to tuning $\gsu(N)$ and $\gsp(N)$ singularities. However, instead of tuning the discriminant to be proportional to some power of a divisor in the base, we tune quantities to be proportional to a power of the $\secz$ component of the section. In non-abelian contexts, models with gauge groups tuned on singular divisors can have non-UFD structure, which can be derived using the normalized intrinsic ring technique discussed in \cite{kmrt}. For the $q=3$ construction, $\secz$ has a singular structure, and the quotient ring $\quotring{\secz}$ is not a UFD. Starting with an ansatz for $\secz$, we can use the normalized intrinsic ring to derive a generalization of the \kmopr model. The intricate structure of the $q=3$ construction is therefore directly linked to the singular nature of $\secz$. Moreover, the normalized intrinsic ring provides a new perspective on the birational equivalence of the $q=3$ and Morrison-Park models.

We first describe the tuning process for abelian models and illustrate the procedure by rederiving the Morrison-Park form. We then briefly review the normalized intrinsic ring technique before using it to derive the $q=3$ construction and analyze its structure. This section concludes with some comments on the matter spectrum and on ways of unHiggsing the $\gu(1)$ symmetry to non-abelian groups. 

\subsection{Tuning abelian models}

For a single $\gu(1)$ group, we need a section $[\secx:\secy:\secz]$ (other than the zero section) such that
\begin{equation}
\secy^2-\secx^3 = \secz^4\left(f\secx+g\secz^2\right). \label{eq:globWeq}
\end{equation}
This expression is simply a rewriting of the global Weierstrass form in \eqref{eq:globWeierstrassform}, with the $x, y, z$ coordinates replaced with components of the section. The left-hand side has a similar structure to the expression for the discriminant $\Delta=4f^3 + 27 g^2$. Moreover, the equation shows that $\secy^2-\secx^3$  must be proportional to $\secz^4$, reminiscent of the conditions for an $I_4$ singularity. These observations suggest that a $\gu(1)$ can be tuned using a method similar to that used for tuning $\gsu(N)$ or $\gsp(N)$ gauge groups:
\begin{enumerate}
\item We first expand $\secx$ and $\secy$ as series in $\secz$. We assume that $\secz$, $\secx$ and $\secy$ are all holomorphic. 
\item We tune $\secx$ and $\secy$ so that
\begin{equation}
\secy^2-\secx^3 \propto \secz^4.\label{eq:u1condition}
\end{equation}
This step bears the most resemblance to the $I_n$ tuning process. 
\item If necessary, we perform additional tunings so that $\secy^2-\secx^3$ is a sum of terms proportional to either $\secz^6$ or $\secx$.
\item Finally, we can read off $f$ and $g$ from the expression for $\secy^2-\secx^3$. 
\end{enumerate}
While the process outlined above is similar to the $I_n$ tuning process, note that, unlike $f$ and $g$ in a standard non-abelian tuning, $\secx$ and $\secy$ can vanish to orders 4 and 6 on some codimension two locus. In fact, this seems to generally happen for $\gu(1)$ models with $q\geq 3$.

To illustrate this procedure, we first consider a situation in which $\secz$ is equal to a generic parameter $b$. We expand $\secx$ and $\secy$ as series in $b$:
\begin{align}
\secx &= x_0 + x_1 b + x_2 b^2 + \ldots & \secy &= y_0 + y_1 b+ y_2 b^2 +\ldots .
\end{align}
Note that we are only interested in expressions for the $x_i$ and $y_i$ up to terms proportional to $b$; for instance, a term proportional to $b$ in $x_i$ can be shifted to $x_{i+1}$ without loss of generality. Said another way, the important properties of $x_i$ and $y_i$ are their images in the quotient ring $\quotring{b}$, in which elements that differ only by terms proportional to $b$ are identified. Here, $R$ refers to the coordinate ring of (an open subset of) the base $B$. Since $b$ is a generic parameter, we assume that $\quotring{b}$ is a unique factorization domain (UFD).

We now need to tune the $x_i$ and $y_i$ so that
\begin{equation}
\secy^2-\secx^3 \propto b^4.
\end{equation}
Plugging the expansions of $\secx$ and $\secy$ gives
\begin{equation}
\left(y_0^2-x_0^3\right) + \left(2 y_0 y_1-3 x_0^2 x_1\right)b + \ldots \propto b^4.
\end{equation}
To perform the tuning, we work order by order, imposing relations such as
\begin{align}
y_0^2-x_0^3 &\equiv 0 \bmod{b},
\end{align}
and so on. Since all of the constraints involve congruence relations modulo $b$, we are essentially considering the conditions to be equations in the quotient ring $\quotring{b}$. But the solutions for $x_i$ and $y_i$ that ensure $\secy^2-\secx^3 \propto b^4$ are already known for situations where $\quotring{b}$ is a UFD. We should use the UFD non-split $I_4$ tuning \cite{tate-f-theory,matter-singularities}, only with the numerical coefficients adjusted:\footnote{The order one terms in the standard $I_4$ tuning can be removed by a redefinition of $\phi$, $x_2$, and $y_4$.}
\begin{align}
\secx &= \phi^2 + x_2 b^2 & \secy &= \phi^3 + \frac{3}{2}\phi x_2 b^2 + y_4 b^4.
\end{align}

These tunings lead to
\begin{equation}
\secy^2-\secx^3 = b^4\left[\left(\phi^2 + x_2 b^2\right)\left(-\frac{3}{4}x_2^2+2 \phi y_4\right)+b^2\left(-\frac{1}{4}x_2^3+x_2 y_4 \phi + y_4^2 b^2\right)\right]
\end{equation}
The right-hand side of this equation already matches the right-hand side of Equation \eqref{eq:u1condition}, so no further tunings are required. We can thus read off that
\begin{align}
f&= -\frac{3}{4}x_2^2+2 \phi y_4+ f_2 b^2 & g&= -\frac{1}{4}x_2^3+x_2 y_4 \phi + y_4^2 b^2 - f_2\left(\phi^2 + x_2 b^2\right)
\end{align}
Notice that we have added and subtracted an $f_2\secx b^2$ term from $\secx^3-\secy^2$, leading to the inclusion of $f_2$ terms in both $f$ and $g$. 

If we redefine parameters as
\begin{align}
x_2 &= -\frac{2}{3}c_2 & \phi &= c_3 & y_4 &= \frac{1}{2}c_1 & f_2 &=-c_0,
\end{align}
we find
\begin{align}
f &= c_1 c_3 -\frac{1}{3}c_2^2-b^2 c_0 & g&= c_0 c_3^2 - \frac{1}{3}c_1c_2 c_3 + \frac{2}{27}c_2^3  -\frac{2}{3}b^2 c_0c_2+\frac{1}{4}b^2c_1^2.
\end{align}
These are exactly the $f$ and $g$ for the Morrison-Park $\gu(1)$ form \cite{morrison-park}. The section, meanwhile, is now given by
\begin{align}
\secx &= c_3^2 -\frac{2}{3}c_2 b^2 & \secy &= c_3^3 - b^2 c_2 c_3+\frac{1}{2}b^4 c_1 & \secz&= b,
\end{align}
which agrees with the expressions in \cite{morrison-park} up to an unimportant negative sign in $\secy$.\footnote{To address the negative sign discrepancy, one can let $b\rightarrow -b$, which changes the sign of $\secz$ but leaves $\secx$ and $\secy$ unchanged. Then, one can scale $(\secx,\secy,\secz)$ by $((-1)^2,(-1)^3,(-1))$ and obtain the exact form of the section in \cite{morrison-park}.}

\subsection{Non-UFD tunings and the normalized intrinsic ring}

Given that the Morrison-Park form seems to arise from the UFD solutions to the tuning conditions, a natural next step is to consider situations in which $\quotring{\secz}$ is not a UFD. In these cases, there are alternative solutions to the tuning constraints, allowing for deviations from the Morrison-Park form. For example, suppose that
\begin{equation}
\secz = \sigma^2 - B \eta^2.
\end{equation}
For this $\secz$, $\quotring{\secz}$ is not a UFD, as explained in more detail below. A constraint such as
\begin{equation}
 \eta y_2 - x_1^2 \equiv 0 \bmod{\secz}, \label{eq:constrexample}
\end{equation}
can be solved in multiple ways. We can let
\begin{align}
x_1 &:= \eta \xi_1 & y_2 &:= \eta\xi_1^2, 
\end{align}
which is a possible solution even if $\quotring{\secz}$ is a UFD. For this solution, $\eta y_2 - x_1^2$ vanishes identically. However, one could also let
\begin{align}
x_1 &:= \sigma & y_2 &:= \eta B.
\end{align}
Then,
\begin{equation}
\eta y_2 - x_1^2 = \eta^2 B - \sigma^2 = -\secz,
\end{equation}
so this second possibility is also a solution. Note that this second solution depends on the specific form of $\secz$, as $\eta y_2-x_1^2$ is an expression that happens to be proportional to the chosen $\secz$.

This example raises two questions: When are multiple solutions possible? And how can we determine the form of the other solutions? Multiple solutions are allowed when $\quotring{\secz}$ is not a UFD and polynomials may have multiple factorizations up to terms proportional to $\secz$. In the example above, $x_1^2$ and $\eta y_2$ represent two distinct ways of factoring the same polynomial in $\quotring{\secz}$, as $x_1^2$ and $\eta y_2$ differ only by a term proportional to $\secz$. As noted in \cite{kmrt}, the quotient ring $R/\mathcal{I}$ for an ideal $\mathcal{I}$ is non-UFD if the variety $V$ corresponding to $\mathcal{I}$ is singular. For the abelian tuning process, we can have a non-UFD $\quotring{\secz}$ if the divisor $\secz = 0$ in the base is singular. This is the case for the \kmopr model: the $\secz$ component is given by
\begin{equation}
\secz =  s_7 s_8^2 - s_6 s_8 s_9 + s_5 s_9^2,
\end{equation}
and the divisor $\secz=0$ has double point singularities at $s_8=s_9=0$. The $q=3$ and $q=4$ models derived here have a singular $\secz$ as well. 

We can obtain the alternative solutions by using the \emph{normalized intrinsic ring} \cite{kmrt}, which we briefly review here. Even if $\secz=0$ is singular, it has a normalization that is smooth in codimension one. The normalized intrinsic ring describes functions on this normalized variety. Consider the ring $\quotring{\secz}$, where $R$ refers to the coordinate ring of (an open subset of) the base $B$. Because the variety $\secz=0$ is singular, $\quotring{\secz}$ is not a UFD. However, the field of fractions of $\quotring{\secz}$ is a UFD. The normalized intrinsic ring, written as $\normring{\secz}$, is defined as the integral closure of this field of fractions, and we can take $\normring{\secz}$ to be a UFD.\footnote{If $\secz$ is one-dimensional (as would be the case for 6D theories), $\normring{\secz}$ is automatically a UFD; see Section 2.4 (particularly Theorem 2.14) of \cite{cutkosky} for further details. In 4D, $\secz=0$ would be complex two-dimensional, and even after normalization there may be singularities at codimension two. Thus, $\normring{\secz}$ may not be a UFD in 4D. To derive the models considered here, we will assume that, regardless of dimension, $\normring{\secz}$ is a UFD.} To construct it explicitly, we add elements from the field of fractions that satisfy a monic polynomial with coefficients in $\quotring{\secz}$. In the $\secz=\sigma^2 - B \eta^2$ example, we know that
\begin{equation}
\left(\frac{\sigma}{\eta}\right)^2 - B = 0.
\end{equation}
We therefore add an element $\tilde{H}$ satisfying $\sigma - \eta \tilde{H} = 0$ and $\tilde{H}^2 = B$. Thus, the normalized intrinsic ring can formally written as
\begin{equation}
\normring{\secz} = R[\tilde{H}]/\langle\sigma - \eta \tilde{H},B-\tilde{H}^2\rangle. \label{eq:normringex}
\end{equation}
We follow the notation in \cite{kmrt}, in which all parameters in the normalized intrinsic ring (that are not well-defined in the quotient ring) are capitalized and marked with a tilde. 

Since we take the normalized intrinsic ring to be a UFD, the solutions to the constraints should be the UFD solutions when we work in the normalized intrinsic ring. For instance, the solution for \eqref{eq:constrexample} would take the form
\begin{align}
x_1 &\sim \eta \tilde{\Xi}_{1} & y_2 &\sim \eta \tilde{\Xi}_{1}^2,
\end{align}
and for simplicity we let $\tilde{\Xi}_{1}$, an element of the normalized intrinsic ring, be $\tilde{H}$. But in the tuning process, $x_1$ and $y_2$ appear in the expansion of the section components, and since we are interested in situations where $\secx$ and $\secy$ are holomorphic, $x_1$ and $y_2$ should be well-defined as elements of $\quotring{\secz}$. We therefore need to use the equivalence relations implied by \eqref{eq:normringex} to remove all instances of $\tilde{H}$. Then,
\begin{align}
\eta \tilde{\Xi}_{1} = \eta\tilde{H} &\rightarrow \sigma & \eta \tilde{\Xi}_{1}^2 = \eta \tilde{H}^2 &\rightarrow \eta B,
\end{align}
and we recover the alternative tuning. In general, finding the non-UFD solutions involves starting with the UFD solutions in the normalized intrinsic ring and determining how to make these expressions well-defined in $\quotring{\secz}$.

\subsection{Tuning models with $q=3$}
We now describe how to systematically derive a $\gu(1)$ construction admitting $q=3$ matter. The goal is to demonstrate that the normalized intrinsic ring techniques can generate $q=3$ models, not to find the most general construction. As such, we will not focus on whether the algebraic tunings used here are the most general possibilities. However, the tuning presented here is more general than the \kmopr construction, as discussed later. 

Our starting point is the assumption that
\begin{equation}
\secz = \ba \dbla^2 + 2 \bb \dbla \dblb + \bc \dblb^2.
\end{equation}
This form for $\secz$ is equivalent to that in the \kmopr model but with differing symbols. Note that the divisor $\secz=0$ in the base would have double point singularities on $\dbla=\dblb=0$, and $\quotring{\secz}$ is not a UFD. The tuning for $\secx$ and $\secy$ can therefore have non-UFD structure, which we derive using the normalized intrinsic ring. For this particular $\secz$, we form the normalized intrinsic ring by adding a new element $\zL$ that satisfies the relations
\begin{gather}
\dblb\zL -  \left(\ba \dbla + \bb \dblb\right) = 0 \label{eq:normringdblb}\\
\dbla\zL +  \left(\bb \dbla + \bc \dblb\right) = 0 \label{eq:normringdbla}\\
\zL^2 - \left(\bb^2 - \ba \bc\right) = 0 .\label{eq:normringsq}
\end{gather}
This normalized intrinsic ring is essentially the same as that used for the symmetric matter models in \cite{kmrt}.

We then expand $\secx$ and $\secy$ as power series in $\secz$.
\begin{align}
\secx &= x_0 + x_1\secz+ x_2 \secz^2 & \secy &= y_0 + y_1 \secz + y_2 \secz^2 + y_3 \secz^3 + y_4 \secz^4.
\end{align}
The series can be truncated at orders $2$ and $4$; if included, higher order terms can be absorbed into other parameters once the tuning is completed. 

For convenience, we define the quantity  $\falsedisc$ to be the left-hand side of \eqref{eq:globWeq}:
\begin{equation}
\falsedisc := \secy^2 - \secx^3.
\end{equation}
In general, we choose notations that agree with the $\gsu(2)$ model discussed in \cite{kmrt}. The symbol $\sim$ indicates that expressions are equivalent when viewed as elements of the normalized intrinsic ring. For instance, an expression such as $x_1 \sim t\zL$ would suggest that $x_1$ is proportional to $\zL$ in the normalized intrinsic ring; however, since $x_1$ should be well-defined in the quotient ring, the expression $t\zL$ must be converted to a well-defined quotient ring expression.

\subsubsection{Canceling terms up to fourth order}
\paragraph{Order 0 cancellation}

We need
\begin{equation}
y_0^2 - x_0^3 \equiv 0 \bmod{\secz}.
\end{equation}
If $\quotring{\secz}$ were a UFD, the only way to satisfy this constraint would be to have $x_0$ and $y_0$ be proportional to the square and cube of some parameter, respectively. This parameter is the equivalent of the $c_3$ parameter in the Morrison Park tuning. For the case at hand, $\quotring{\secz}$ is not a UFD, but $\normring{\secz}$ is a UFD. In principle, we can therefore let $x_0$ and $y_0$ be proportional to the square and cube of some parameter $\tilde{T}$ in  $\normring{\secz}$. However, $x_0$ and $y_0$ are elements of the coordinate ring and must have  well-defined expressions in $\quotring{\secz}$. In fact, for the $\secz$ considered here, $\tilde{T}^2$ and $\tilde{T}^3$ are well-defined in $\quotring{\secz}$ only if $\tilde{T}$ is well-defined in $\quotring{\secz}$.\footnote{See Section 5 of \cite{kmrt} for a more detailed discussion.} Thus, we can set
\begin{align}
x_0 &:= \cthree^2 & y_0 &:= \cthree^3,
\end{align}
where $\cthree$ is well-defined in $\quotring{\secz}$. With these definitions, $y_0^2-x_0^3$ vanishes identically, and $\falsedisc$ is proportional to $\secz$. 

\paragraph{Order 1 cancellation}
The condition for $\falsedisc \propto \secz^2$ is that
\begin{equation}
\cthree^3\left(2 y_1 - 3 \cthree x_1\right) \equiv 0 \bmod{\secz}.
\end{equation}
This condition can be satisfied by setting
\begin{equation}
y_1 := \frac{3}{2}\cthree x_1. 
\end{equation}
$\falsedisc$ is now proportional to $\secz^2$.

\paragraph{Order 2 cancellation}
The condition for $\falsedisc \propto \secz^3$ is that
\begin{equation}
\cthree^2\left(2\cthree y_2 -\frac{3}{4}x_1^2-3\cthree^2 x_2\right) \equiv 0 \bmod{\secz}.
\end{equation}
If we work in $\normring{\secz}$, which is a UFD, the only way to satisfy this condition (without forcing $\cthree$ to be a perfect square) is to have
\begin{align}
x_1 &\sim \frac{1}{6}\cthree \PsiL & y_2 &\sim  \frac{3}{8}\cthree \left(\frac{1}{36}\PsiL^2 + 4 x_2\right). \label{eq:x1y2normdef}
\end{align}
$\PsiL$ is an element of $\normring{\secz}$, which we can write as \footnote{One could use the more general expression $\PsiL = \xi_{(1)}\dbla + \xi_{(0)} \dblb + \phibar \zL$. However, after the full tuning is completed, $\xi_{(0)}$ and $\xi_{(1)}$ can be removed by redefinitions of the other parameters in the Weierstrass model. We therefore drop $\xi_{(0)}$ and $\xi_{(1)}$ from the beginning to simplify the discussion.}
\begin{equation}
\PsiL = \phibar \zL
\end{equation}

However, $x_1$ and $y_2$ are elements of the coordinate ring, and the above tunings involving $\zL$ must be rewritten as expressions that are well-defined in $\quotring{\secz}$. To obtain a non-trivial tuning, we should not tune $\phibar$ in a way that makes $\PsiL$ well-defined in $\quotring{\secz}$. Therefore, in order for both $x_1$ and $\cthree$ to be well-defined, $\cthree$ must take the form
\begin{equation}
t :=  \thetaa \dbla + \thetab \dblb.
\end{equation}
Using \eqref{eq:normringdblb} and \eqref{eq:normringdbla} to replace $\zL \dblb$ and $\zL\dbla$ with expressions in $\quotring{\secz}$, we define $x_1$ to be
\begin{equation}
x_1 := \frac{1}{6}\phibar\left[\thetab \left(\ba \dbla + \bb \dblb\right) - \thetaa\left(\bb \dbla + \bc \dblb\right)\right].
\end{equation}
Meanwhile, \eqref{eq:normringsq} implies that $y_2$ should be defined to be
\begin{equation}
y_2 := \frac{3}{8} t \left[\frac{1}{36}\phibar^2 \left(\bb^2 - \ba \bc\right)+4 x_2\right]
\end{equation}

With these tunings,
\begin{equation}
\cthree^2\left(2\cthree y_2 -\frac{3}{4}x_1^2-3\cthree^2 x_2\right)  = -\frac{1}{48}\cthree^2 \phibar^2\left(\ba \thetab^2 - 2 \bb \thetab \thetaa+ \bc\thetaa^2\right)\secz,
\end{equation}
and $\falsedisc$ is proportional to $\secz^3$. For convenience, we define the quantity $\ordertworem$ to be
\begin{equation}
\ordertworem := -\frac{1}{48}\cthree^2 \phibar^2\left(\ba \thetab^2 - 2 \bb \thetab \thetaa+ \bc \thetaa^2\right).
\end{equation}

\paragraph{Order 3 cancellation} The condition for $\falsedisc \propto \secz^4$ is that
\begin{equation}
\ordertworem - x_1^3 + 3 \cthree x_1 \left(y_2 - 2\cthree x_2\right) + 2\cthree^3 y_3 \equiv 0 \bmod{\secz}.
\end{equation}
In $\normring{\secz}$, this condition can be written as
\begin{multline}
\cthree^2\Bigg[-\frac{1}{48}\phibar^2 \left(\ba \thetab^2 - 2 \bb \thetab \thetaa+ \bc \thetaa^2\right)\\
+\cthree\left(2 y_3+\frac{1}{1728}\PsiL^3 -\frac{1}{4} x_2 \PsiL\right)\Bigg] = 0.
\end{multline}
The contributions from $\ordertworem$ cannot be canceled without further tunings: for instance, the other terms within the square brackets are proportional to either $\dbla$ or $\dblb$, while the contributions from $\ordertworem$ are not. We should not use tunings that change the form of $\secz$ or tune $\phibar$ in a way that removes the non-UFD structure. But we can introduce $\dbla$ and $\dblb$ factors by tuning $\thetaa$ and $\thetab$.  In particular, we can let
\begin{align}
\thetaa &= \taua \dbla + \taub \dblb & \thetab &= \taub^\prime \dbla + \tauc \dblb.
\end{align}
Additionally, $\ba \thetab^2 - 2 \bb \thetab \thetaa+ \bc^2 \thetaa^2$ should be the sum of two terms: one proportional to $\cthree$, and the other proportional to $\secz$. This is not the case after the tunings done so far, but we can satisfy this condition by letting $\taub = \taub^\prime$. We therefore define $\thetab$ and $\thetaa$ as
\begin{align}
\thetaa &:= \taua \dbla + \taub \dblb & \thetab &:= \taub \dbla + \tauc \dblb,
\end{align}
and $\cthree$ is quadratic in $\dbla$ and $\dblb$:
\begin{equation}
\cthree = \taua \dbla^2 + 2 \taub \dbla\dblb + \tauc \dblb^2.
\end{equation}
Now, 
\begin{equation}
\ba \thetab^2 - 2 \bb \thetab \thetaa+ \bc \thetaa^2 = \left(\taub^2 -\taua \tauc\right)\secz + \left(\ba \tauc - 2 \bb \taub + \bc \taua\right)\cthree.
\end{equation}
The $\ordertworem$ terms can now be canceled by letting
\begin{equation}
y_3 := y_3^\prime + \frac{1}{96}\phibar^2\left(\ba \tauc - 2 \bb \taub + \bc \taua\right),
\end{equation}
at least up to terms proportional to $\secz$.

The third order cancellation condition now reads
\begin{equation}
\cthree^3\left(2 y_3^\prime+\frac{1}{1728}\PsiL^3 -\frac{1}{4} x_2 \PsiL\right)=0
\end{equation}
$\PsiL^3$ is not well defined in $\quotring{\secz}$, so we cannot use $y_3^\prime$ to cancel this term. But working in $\normring{\secz}$, we can cancel the remaining terms using tunings that, in $\normring{\secz}$, take the form
\begin{align}
x_2 &\sim -\frac{1}{6}\ctwo +\frac{1}{432}\PsiL^2 & y_3^\prime &\sim -\frac{1}{48}\ctwo \PsiL.
\end{align}
We can immediately convert the $x_2$ expression into a well-defined quantity in $\quotring{\secz}$, giving the following definition for $x_2$:
\begin{equation}
x_2 := -\frac{1}{6}\ctwo + \frac{1}{432}\phibar^2\left(\bb^2 - \ba \bc\right)
\end{equation}
The $\ctwo \PsiL$ term in the $y_3^\prime$ expression, however, cannot be written in $\quotring{\secz}$ without further tuning $\ctwo$. $\ctwo$ must be well-defined in $\quotring{\secz}$, so it should take the form
\begin{equation}
\ctwo := \phia \dbla + \phib \dblb.
\end{equation}
Then, $y_3^\prime$ should be defined as
\begin{equation}
y_3^\prime := -\frac{1}{48}\phibar\left[\phib\left(\ba \dbla+\bb \dblb\right)-\phia \left(\bb \dbla + \bc \dblb\right)\right].
\end{equation}

To summarize, we have performed the following tunings:
\begin{align}
\thetaa :=& \taua \dbla + \taub \dblb\\
\thetab :=& \taub \dbla + \tauc \dblb\\
x_2 :=& -\frac{1}{6}\left(\phia \dbla + \phib \dblb\right) + \frac{1}{432}\phibar^2\left(\bb^2 - \ba \bc\right)\\
y_3 :=& \frac{1}{96}\phibar^2\left(\ba \tauc - 2 \bb \taub + \bc \taua\right)\notag\\
&-\frac{1}{48}\phibar\left[\phib\left(\ba \dbla+\bb \dblb\right)-\phia \left(\bb \dbla + \bc \dblb\right)\right].
\end{align}
With these tunings,
\begin{equation}
\ordertworem - x_1^3 + 3 \cthree x_1 \left(y_2 - 2\cthree x_2\right) + 2\cthree^3 y_3= \orderthreerem \secz,
\end{equation}
where
\begin{multline}
\orderthreerem= -\frac{1}{48}\cthree^2 \phibar^2 \left(\taub^2 - \taua \tauc\right) + \frac{1}{24}\cthree^2 \phibar\left[\phia\left(\taub \dbla + \tauc \dblb\right)-\phib\left(\taua \dbla + \taub \dblb\right)\right]\\
-\frac{1}{36}\phibar^2 x_1
\left[\left(\taub^2 -\taua \tauc\right)\secz + \left(\ba \tauc - 2 \bb \taub + \bc \taua\right)\cthree\right]\label{eq:order3rem}
\end{multline}
and
\begin{equation}
\cthree = \taua \dbla^2 + 2 \taub \dbla \dblb + \tauc \dblb^2.
\end{equation}
$\falsedisc$ is therefore proportional to $\secz^4$.

\subsubsection{Finding $f$ and $g$}

Ultimately, we need to extract $f$ and $g$ from the relation
\begin{equation}
\secy^2 - \secx^3 = \secz^4\left(f \secx+g \secz^2\right)
\end{equation}
Now that $\secy^2 - \secx^3$ is proportional to $\secz^4$, we can start extracting portions of $f$ and $g$. Unlike in the Morrison-Park case, we need to further tune parameters in $\secx$ and $\secy$ to extract $f$ and $g$. 

As in the previous steps, we will work order by order. If we let
\begin{equation}
f = f_0 + f_1 \secz + f_2 \secz^2,
\end{equation}
we have the condition that
\begin{equation}
\secy^2 - \secx^3 - \cthree^2 f_0\secz^4 - \left(\cthree^2 f_1 + x_1 f_0\right)\secz^5 \propto \secz^6 . \label{eq:order45eq}
\end{equation}
Our goal is now to cancel the order 4 and order 5 terms on the left-hand side of the above equation.

\paragraph{Order 4 Cancellation}
The condition for the order 4 terms to cancel is that
\begin{equation}
\orderthreerem + y_2^2 - 3 x_1^2 x_2+3 \cthree x_1 y_3 - 3 \cthree^2 x_2^2+2 \cthree^3 y_4- \cthree^2 f_0 \equiv 0 \bmod{\secz}, \label{eq:ord4cancond}
\end{equation}
where $\orderthreerem$ is given by \eqref{eq:order3rem}. There are several terms in the above expression that are explicitly proportional to $\cthree^2$. Such terms can fairly easily be canceled by tuning $f_0$ to take the form
\begin{multline}
f_0 := -\frac{1}{48}\left(\dblb \phib + \dbla \phia\right)^2+\frac{1}{24}\phibar\left[\phia\left(\taub \dbla + \tauc \dblb\right)-\phib\left(\taua \dbla + \taub \dblb\right)\right]\\
-\frac{1}{48} \phibar^2 \left(\taub^2 - \taua \tauc\right) + 2 \cthree y_4 - \frac{\phibar^2}{216}\left(\bb^2-\bc\ba\right)\left(\dblb \phib + \dbla \phia\right)\\
+\frac{11}{48}\frac{1}{1296}\phibar^4\left(\bb^2-\bc\ba\right)^2 + f_0^\prime.
\end{multline}
The cancellation condition now takes the form
\begin{multline}
\cthree x_1\left[-\frac{1}{16}\phibar\left(\phib\left(\bb\dblb+\ba\dbla\right)-\phia\left(\bc\dblb+\bb\dbla\right)\right)+\frac{3}{864}\phibar^2\left(\ba\tauc-2\bb\taub+\bc\taua\right)\right]\\
+x_1^2 \left[\frac{1}{2}\left(\dblb \phib + \dbla \phia\right)-\frac{1}{144}\phibar^2\left(\bb^2-\bc\ba\right)\right] - f_0^\prime \cthree^2 \equiv 0 \bmod{\secz}.
\end{multline}
Working in $\normring{\secz}$, this condition is equivalent to
\begin{multline}
\cthree^2\Bigg[-\frac{\phibar^4}{5184}\zL^4+\frac{\phibar^2}{288} \zL^2 \left(\phib \dblb+\phia \dbla\right)-f_0^\prime\\
+\frac{\phibar^3}{1728}\zL\left(\ba\tauc-2\bb\taub+\bc\taua\right)\Bigg]=0 \label{eq:f0primenir}
\end{multline}
If all the terms in square brackets were well-defined in $\quotring{\secz}$, we could immediately read off an expression for $f_0^\prime$ that would cancel terms. However, this is not currently the case. The terms that have even powers of $\zL$ are already well-defined in $\quotring{\secz}$, since $\zL^2$ is equivalent to $\bb^2-\bc\ba$. But the $\zL$ term in the square brackets is currently not well-defined in $\quotring{\secz}$. Without modifying $\phibar$, which would lead to a trivial tuning, the only way to fix this term is to force $(\ba\tauc-2\bb\taub+\bb\taua)$ to be a sum of terms proportional to $\dbla$ or $\dblb$. This can be accomplished with the ansatz that $\tauc$, $\taub$, and $\taua$ take the form
\begin{align}
\tauc &:= \td \dblb+\tc\dbla& \taub &:= \tc \dblb + \tb \dbla & \taua &:= \tb \dblb + \ta \dbla.\label{eq:tauredef}
\end{align}
These tunings make $\cthree$ cubic in $\dbla$ and $\dblb$, as
\begin{equation}
\cthree  = \ta \dbla^3 + 3 \tb \dbla^2 \dblb + 3 \tc\dbla \dblb^2 + \td \dblb^3.
\end{equation}
Now, the third term in \eqref{eq:f0primenir} is well-defined in $\quotring{\secz}$, as
\begin{multline}
\zL\left(\ba\tauc-2\bb\taub+\bc\taua\right) = \left(\ba \td - 2 \bb \tc+\bc \tb\right)\left(\ba\dbla+\bb\dblb\right)\\-\left(\ba \tc-2\bb\tb+\bc\ta\right)\left(\bb \dbla+\bc\dblb\right)
\end{multline}
in $\normring{\secz}$. We thus define $f_0^\prime$ to be
\begin{multline}
f_0^\prime = -\frac{1}{5184}\phibar^4\left(\bb^2 - \ba \bc\right)^2+\frac{1}{288}\phibar^2\left(\bb^2 - \ba \bc\right)\left(\phib \dblb+\phia\dbla\right)\\
+\frac{1}{1728}\phibar^3\Bigg[\left(\ba \td - 2 \bb \tc+\bc \tb\right)\left(\ba\dbla+\bb\dblb\right)\\-\left(\ba \tc-2\bb\tb+\bc\ta\right)\left(\bb \dbla+\bc\dblb\right)\Bigg].
\end{multline}
The left-hand side of \eqref{eq:ord4cancond} is now equal to $\orderfourrem \secz$, where
\begin{multline}
\orderfourrem = -\frac{1}{16}\phibar x_1\left[\phib\left(\taub\dblb+\taua\dbla\right)-\phia\left(\tauc\dblb+\taub\dbla\right)\right]-\frac{1}{48}\phibar^2 x_1 \tausq + \frac{1}{864}\phibar^3 \taucu \secz \\
+\frac{1}{288}\phibar^2\left(\phib \dblb+\phia \dbla-\frac{\phibar^2}{18}\left(\bb^2-\bc\ba\right)\right)\times\\
\left[\left(\ba \tauc - 2 \bb \taub + \bc \taua\right)\cthree + \tausq\secz\right].
\end{multline}
$\tausq$ is given by $\taub^2-\tauc\taua$, and
\begin{multline}
\taucu = \frac{1}{2}\Bigg[-\left(2\tb^3-3 \tc\tb\ta+\td\ta^2\right)\dbla^3-3\left(\tc \tb^2-2\tc^2\ta+\td\tb\ta\right)\dbla^2\dblb\\
+3\left(\tc^2\tb - 2 \td\tb^2+\td \tc \ta\right)\dbla \dblb^2+\left(2\tc^3-3 \td \tc \tb + \td^2 \ta\right)\dblb^3\Bigg].
\end{multline}
$\secy^2-\secx^3 - f \secx \secz^4$ is therefore proportional to $\secz^5$. 

To summarize, we have tuned $\taua$, $\taub$, and $\tauc$ to take the form in Equation \eqref{eq:tauredef} and have found $f_0$ to be
\begin{multline}
f_0 := -\frac{1}{48}\left(\dblb \phib + \dbla \phia\right)^2+ \frac{1}{24}\phibar\left[\phia\left(\taub \dbla + \tauc \dblb\right)-\phib\left(\taua \dbla + \taub \dblb\right)\right]\\
-\frac{1}{48} \phibar^2 \left(\taub^2 - \taua \tauc\right) + 2 \cthree y_4 \\
-\frac{1}{864} \phibar^2\left(\bb^2-\bc\ba\right)\left(\dblb \phib + \dbla \phia\right)
-\frac{1}{62208}\phibar^4\left(\bb^2-\bc\ba\right)^2\\
+ \frac{1}{1728}\phibar^3\Bigg[\left(\ba \td - 2 \bb \tc+\bc \tb\right)\left(\ba\dbla+\bb\dblb\right)\\-\left(\ba \tc-2\bb\tb+\bc\ta\right)\left(\bb \dbla+\bc\dblb\right)\Bigg].
\end{multline}

 \paragraph{Order 5 Cancellation} The condition for the order $\secz^5$ terms to cancel is that
 \begin{equation}
 \orderfourrem - 3 x_1 x_2^2+ 2 y_2 y_3 +2 y_1 y_4 - f_0 x_1 - f_1 x_0 \equiv 0 \bmod{\secz}.
 \end{equation}
 Using the previous expressions for the various parameters, this can be rewritten as
\begin{multline}
-\cthree^2 \fonesec + x_1\left[ \cthree y_4 - \frac{1}{48}\phibar\left(\phib\left(\taua \dbla+\taub\dblb\right)-\phia\left(\taub \dbla+\tauc \dblb\right)\right)\right]\\
 -\frac{1}{576}\phibar^2 \cthree \left(\phia \dbla+\phib \dblb\right) \left(\ba \tauc - 2 \bb \taub + \bc \taua\right) \equiv 0 \bmod{\secz}.
\end{multline}

Working in $\normring{\secz}$, the cancellation condition reads
\begin{multline}
\cthree^2\left[\frac{1}{6}\phibar \zL y_4 -\fonesec \right]-\frac{1}{576}\phibar^2\cthree\Bigg[2\tilde{B}\left(\phib\left(\taua \dbla+\taub\dblb\right)-\phia\left(\taub \dbla+\tauc \dblb\right)\right) \\+ \left(\phia \dbla+\phib \dblb\right) \left(\ba \tauc - 2 \bb \taub + \bc \taua\right)\Bigg] = 0.
\end{multline}
The $\cthree^2$ term is order 6 in $\dbla,\dblb$, as $\cthree$ is order 3 in $\dbla,\dblb$.\footnote{Note that converting expressions involving $\tilde{B}$ to well-defined expressions in $\quotring{\secz}$ does not change the order of the expression in $\dbla$, $\dblb$. } However, the other terms are order 5 in $\dbla,\dblb$. (Recall that $\tauc$, $\taub$, and $\taua$ are all order 1 in $\dbla,\dblb$, as can be seen from \eqref{eq:tauredef}.) These terms can be canceled only if we perform some tuning to increase their order in $\dbla,\dblb$. Making $\td$ through $\ta$ proportional to $\dbla$ and $\dblb$ will not fix the issue; this tuning would increase the orders of both $\cthree$ and the order 5 terms, and the mismatch in orders would persist. But we can tune $\phib$ and $\phia$ to be \footnote{One could consider a more general redefinition $\phia = \ha \dbla + \hb \dblb$, $\phia = \ha \dbla + \hb^\prime \dblb$. However, by performing shifts in the other parameters (namely $y_4$), one can set $\hb=\hb^\prime$ without loss of generality.}
\begin{align}
\phia &:= \ha \dbla + \hb \dblb & \phib &:= \hb \dbla + \hc \dblb.
\end{align}
With these redefinitions, the $\normring{\secz}$ cancellation condition becomes (after dropping terms proportional to $\secz$)
\begin{equation}
\cthree^2\left[\frac{1}{6}\phibar \zL y_4 -\frac{1}{576}\phibar^2\left(\ha\bc-2\hb\bb+\hc\ba\right)-\fonesec\right] = 0.
\end{equation}

If the remaining terms were all well-defined in $\quotring{\secz}$, we could immediately read off the $\fonesec$ tuning that would cancel the remaining terms. However, the $\zL y_4$ is currently ill-defined as an element of $\quotring{\secz}$. $y_4$ must therefore be written as a sum of terms proportional to $\dbla$ and $\dblb$:
\begin{equation}
y_4 := \frac{1}{2}\left(\lambdaa \dbla + \lambdab \dblb\right).
\end{equation}
We can now tune $\fonesec$ to cancel all of the order $\secz^5$ terms:
\begin{multline}
\fonesec = - \frac{1}{576}\phibar^2\left(\ha\bc-2\hb\bb+\hc\ba\right) \\
+ \frac{1}{12}\phibar\left[\lambdab\left(\ba \dbla + \bb \dblb\right)-\lambdaa\left(\bb \dbla+\bc \dblb\right)\right].
\end{multline}

Finally,
\begin{equation}
\secy^2 - \secx^3 - f \secx \secz^4 \propto \secz^6,
\end{equation}
and the tuning process is complete. 

\subsection{Structure of the charge-3 construction}
The $f$, $g$ and section components for the $q=3$ model are given in Appendix \ref{app:charge3appendix}. The homology classes of the various parameters, which are listed in Table \ref{tab:q3params}, can be found by requiring that $f$ and $g$ are respectively sections of $-4K_B$ and $-6K_B$, where $K_B$ is the canonical class of the base. 

\begin{table}
\begin{center}
\begin{tabular}{|c|c|c|}\hline
Parameter & Homology Class & Equivalent in \cite{kmpopr}\\\hline
$\dbla$ & $[\dbla]$& $-s_8$\\
$\dblb$ & $[\dblb]$& $s_9$\\
$\bc$ & $[\secz]-2[\dblb]$& $s_5$\\
$\bb$ & $[\secz]-[\dbla]-[\dblb]$& $s_6/2$\\
$\ba$ & $[\secz]-2[\dbla]$& $s_7$\\
$\phibar$ & $-K_B - [\secz]+[\dbla]+[\dblb]$& $-12$ \\
$\td$ & $-K_B + [\secz]-3[\dblb]$& $s_1$\\
$\tc$ & $-K_B + [\secz]-[\dbla]-2[\dblb]$& $s_2/3$ \\
$\tb$ & $-K_B + [\secz]-2[\dbla]-[\dblb]$& $s_3/3$\\
$\ta$ & $-K_B + [\secz]-3[\dbla]$& $s_4$\\
$\hc$ & $-2 K_B - 2[\dblb]$& $0$\\
$\hb$ & $-2 K_B - [\dbla]-[\dblb]$& $0$\\
$\ha$ & $-2 K_B - 2[\dbla] $ & $0$\\
$\lambdab$ & $-3K_B - [\secz]-[\dblb]$&$0$ \\
$\lambdaa$ & $-3K_B - [\secz]-[\dbla]$& $0$ \\
$f_2$ & $-4 K_B - 2[\secz]$& $0$\\\hline
\end{tabular}
\end{center}
\caption{Parameters for the $q=3$ model. The center column lists the homology classes of the parameters in terms of the homology classes for $\dbla$, $\dblb$, and $\secz$. $K_B$ refers to the canonical class of the base. The rightmost column gives the dictionary between the parameters used here and those used in the previous model in \cite{kmpopr}.}
\label{tab:q3params}
\end{table}

Even though the $q=3$ model differs from the Morrison-Park form, there is a link between the two models. \cite{morrison-park-2016} pointed out that the \kmopr Weierstrass model is birationally equivalent to one in Morrison-Park form, but the Morrison-Park form model may not satisfy Calabi-Yau condition. A similar phenomenon occurs for the $q=3$ construction derived here. If we allow division by $\dblb$, the $q=3$ Weierstrass model can in fact be written in the Morrison-Park form
\begin{align}
f&= \mpa_1 \mpa_3 -\frac{1}{3} \mpa_2^2 - b^2 \mpa_0 & g&=\mpa_0 \mpa_3^2-\frac{1}{3}\mpa_1 \mpa_2 \mpa_3+\frac{2}{27}\mpa_2^3 -\frac{2}{3}b^2 \mpa_0 \mpa_2 + \frac{1}{4}b^2 \mpa_1^2,
\end{align}
with
\begin{align}
\mpa_3 =& -\left(\cthree+\frac{\phibar}{12}\frac{\ba\dbla+\bb\dblb}{\dblb}\secz\right),\label{eq:mpa3}\\
\mpa_2 =& \frac{1}{4}\left(\ha\dbla^2+2\hb\dbla\dblb+\hc\dblb^2 + \frac{\phibar^2}{36}\left(\bb^2-\bc\ba\right)\right)\notag\\
&+\frac{\phibar}{4}\frac{\ta\dbla^2+2\tb\dbla\dbla+\tc\dblb^2}{\dblb}+\frac{\ba\phibar^2}{96\dblb^2}\secz,\\
\mpa_1 =& -\Bigg(\lambdaa \dbla+\lambdab\dblb+\frac{\phibar}{24}\frac{\ha\dbla+\hb\dblb}{\dblb}\notag\\
&\qquad+\frac{\phibar^2}{48}\frac{\ta\dbla+\tb\dblb}{\dblb^2}+\frac{\ba\phibar^3}{1728}\frac{\ba\dbla+\bb\dblb}{\dblb^3}\Bigg),\\
\mpa_0 =& -f_2 +\frac{\phibar}{12}\frac{\lambdaa}{\dblb}+\frac{\phibar^2}{576}\frac{\ha}{\dblb^2}+\frac{\phibar^3}{1728}\frac{\ta}{\dblb^3}+\frac{\phibar^4}{82944}\frac{\ba^2}{\dblb^4},\\
b =& \secz .\label{eq:mpab}
\end{align}
Since the $q=3$ tuning was derived using the normalized intrinsic ring, this observation comes as no surprise. Recall that $\zL$ is in the field of fractions of $\quotring{\secz}$, and for the tuning, we use the UFD structures but include a dependence on $\zL$. The normalized intrinsic ring essentially provides a convenient method for determining how the $\mpa_i$ parameters can depend on fractional terms so that all the fractional terms cancel when $f$ and $g$ are expanded. Indeed, the expressions for the $\mpa_i$ involve $(\ba\dbla+\bb\dblb)/\dblb$, which, in the field of fractions, is equivalent to $\zL$. But the expressions in \eqref{eq:mpa3} through \eqref{eq:mpab} also imply that $\dblb^4 f$ and $\dblb^6 g$ can be written in Morrison-Park form \emph{without} division by $\dblb$: 
\begin{align}
\dblb^4f&= c_1 c_3 -\frac{1}{3} c_2^2 - b^2 c_0 & \dblb^6g&=c_0 c_3^2-\frac{1}{3}c_1 c_2 c_3+\frac{2}{27}c_2^3 -\frac{2}{3}b^2 c_0 c_2 + \frac{1}{4}b^2 c_1^2,
\end{align}
with
\begin{align}
c_3 &= \dblb \mpa_3 & c_2 &= \dblb^2 \mpa_2 & c_1 &= \dblb^3\mpa_1 & c_0 &= \dblb^4 \mpa_0.
\end{align}
In other words,
\begin{equation}
{y^{\prime}}^2 = {x^\prime}^3 + \dblb^4 f x^\prime + \dblb^6 g \label{eq:blowup-charge3}
\end{equation}
is a bona-fide Weierstrass model in Morrison-Park form. This new Weierstrass model is a non-minimal transformation of the $q=3$ model: if $f\in -4K_B$ and $g\in -6K_B$, then $\dblb^4 f\in -4K_B+4[\dblb]$ and $g\in -6K_B+6[\dblb]$. Unless $[\dblb]$ is trivial (in which case there is no $q=3$ matter), the Morrison-Park form Weierstrass model will not be Calabi-Yau. Thus, we see that the $q=3$ model is birationally equivalent to the Morrison-Park form, with the Morrison-Park model satisfying the Calabi-Yau condition only when $q=3$ matter is not present. This is in agreement with the results of \cite{morrison-park-2016}. 

In some sense, the normalized intrinsic ring led to the specific tunings of the $\mpa_i$ that allow the Morrison-Park form model to be blown down to the $q=3$ model, even though we did not use the normalized intrinsic ring directly in this fashion. One might therefore be tempted to use the following strategy to obtain this $q=3$ construction or even other models: start with the Morrison-Park form, let the parameters be rational in, say, $\dblb$, and determine the appropriate expressions that allow the fractional terms to cancel. While this strategy may indeed work, determining the exact structures that enable the correct cancellations may be challenging. For instance, in the construction presented here, $\cthree$ has a cubic structure, and it at least naively seems difficult to predict the particular form that $\cthree$ must take without the help of the tuning procedure. Of course, this alternative strategy may prove fruitful for obtaining new models and would be interesting to explore further. 

Finally, we note that the Weierstrass model we have derived is a generalization of the \kmopr construction. In particular, we can recover the previous $q=3$ construction by setting various parameters to particular values. The dictionary between the parameters used here and those used in the \kmopr model is given in Table \ref{tab:q3params}. Note that we must set $\phibar$ to a constant in order to recover the \kmopr model, forcing a relation between the unspecified homology classes in Table \ref{tab:q3params}:
\begin{equation}
[\secz] = -K_B +[\dbla]+[\dblb].
\end{equation}
The tuning derived here can therefore produce a wider variety of models. For example, suppose we take our compactification base to be $\mathbb{P}^2$ and consider the situation with $[\dbla]=[\dblb]=H$. As discussed shortly, this is a situation with a single $q=3$ hypermultiplet in six dimensions. The \kmopr model requires that $[\secz]=5H$, whereas $[\secz]$ is not restricted to a single homology class in the model derived here. In turn, the new $q=3$ construction admits a wider range of matter spectra.

\subsection{Matter spectra}
\label{subsec:q3matter}
The $q=3$ model has several codimension two $I_2$ loci that support charged matter. In general, $I_2$ loci occur where $\secy = 3\secx^2+f \secz^4 = 0$, but this locus consists of several sub-loci supporting different types of charged matter. We therefore need to examine the expression further to determine the loci corresponding to particular charges. The types of charges supported and the corresponding $I_2$ loci are summarized in Table \ref{tab:q3loci}, and their multiplicities are given in Table \ref{tab:q3matter}. Our matter spectrum analysis will focus primarily on 6D models.  

\begin{table}
\begin{center}
\begin{tabular}{|c|c|}\hline
Charge & $I_2$ locus  \\\hline
$3$ & $V(I_{q=3}) := \{\dbla = \dblb = 0\}$ \\
$2$ & $V(I_{q=2}) := \{\cthree = \secz = 0\}/V(I_{q=3})$ \\
$1$ & $V(I_{q=1}) := \{\secy = f\secz^4 + 3\secx^2 = 0\}/\left(V(I_{q=3})\cup V(I_{q=2})\right)$\\\hline
\end{tabular}
\end{center}
\caption{Charged matter loci for the $q=3$ model.}
\label{tab:q3loci}
\end{table}

\begin{table}
\begin{center}
\begin{tabular}{|c|c|}\hline
Charge &  Multiplicity \\\hline
$3$ &  $\mthree=[\dbla]\cdot[\dblb]$\\
$2$ &  $\mtwo=(-K_B + [\secz])\cdot[\secz]-6[\dbla]\cdot[\dblb]$ \\
$1$ &  $\mone = 12(-K_B +[\secz])\cdot(-K_B +[\secz]) - 81 \mthree - 16\mtwo $\\\hline
\end{tabular}
\end{center}
\caption{Charged matter multiplicities for the $q=3$ model.}
\label{tab:q3matter}
\end{table}

From the dictionary relating the $q=3$ model derived here to the \kmopr model, we know that the $\dbla=\dblb=0$ locus supports $q=3$ matter and that the $q=3$ multiplicity is $[\dbla]\cdot[\dblb]$. $q=2$ matter occurs at loci, apart from the $q=3$ locus, at which all of the section components vanish. Importantly,
\begin{align}
\secx &\equiv \cthree^2 \bmod{\secz} & \secx &\equiv \cthree^3 \bmod{\secz}
\end{align}
with $\cthree$ given by 
\begin{equation}
t = \ta \dbla^3 + 3 \tb \dbla^2 \dblb + 3 \tc \dbla \dblb^2 + \td \dblb^3.
\end{equation}
This implies that the section components vanish at loci where $\secz=\cthree=0$. However $\dbla=\dblb=0$ is a solution to $\secz=\cthree=0$, so we must exclude the $\dbla=\dblb=0$ locus from the $q=2$ locus. This leads us to describe the $q=2$ locus as
\begin{equation}
V(I_{q=2})= \{\cthree = \secz = 0\}/V(I_{q=3})
\end{equation}
where $V(I_{q=3})$ is the variety corresponding to the ideal $\dbla=\dblb=0$. This result is in exact agreement with the $q=2$ locus of the \kmopr model \cite{kmpopr}. To count the $q=2$ multiplicity, we must find the multiplicity of $\dbla=\dblb=0$ within $\secz=\cthree=0$. Here, we use the resultant method described in \cite{cvetic-klevers-piragua}. The resultant of $\cthree$ and $\secz$ with respect to $\dbla$ is given by
\begin{equation}
\Res_{\dbla}(\cthree,\secz) = \dblb^6\rtwo,
\end{equation}
where $\rtwo$ is a long expression independent of $\dblb$. The $\dblb^6$ factor in the resultant indicates that $\dbla=\dblb=0$ has multiplicity 6 within $\secz=\cthree=0$. The $q=2$ multiplicity is therefore given by
\begin{equation}
\mtwo = [\secz]\cdot[\cthree]-6[\dbla]\cdot[\dblb] = [\secz]\cdot(-K_B + [\secz])-6[\dbla]\cdot[\dblb].
\end{equation}
This expression exactly matches the anomaly equation \eqref{eq:abeliangenus}, with the multiplicity $6$ of the $\dbla=\dblb=0$ locus corresponding to the $q^2(q^2-1)/12$ factor in the anomaly equation. \eqref{eq:abeliangenus} therefore seems to describe the loci at which all of the components of the section vanish. In fact, a similar phenomenon occurs in the $q=4$ model described later, hinting that \eqref{eq:abeliangenus} may have a general F-theory interpretation.

$q=1$ matter occurs at the $\secy = 3\secx^2+f \secz^4 = 0$ loci that do not support $q=2$ or $q=3$ matter. The $q=1$ locus can therefore be written as
\begin{equation}
V(I_{q=1}) = \{\secy = 3\secx^2+f \secz^4 =0\}/\left(V(I_{q=3})\cup V(I_{q=2})\right).
\end{equation}
To determine the $q=1$ multiplicity, we must find the multiplicities of $V(I_{q=3})$ and $V(I_{q=2})$ within $\secy = 3\secx^2+f \secz^4 =0$. Again, this information can be read off from the resultant, but evaluating the resultant in this case is computationally intensive. We therefore calculate the resultant in situations where all parameters except $\dbla$, $\dblb$, and $\bc$ are set to random integers. Regardless of the specific integers chosen, the resultant factorizes into the form
\begin{equation}
\Res_{\dbla}(\secy, 3\secx^2+f\secz^4) = \dblb^{81}\rtwo^{16}\rone,
\end{equation}
where $\rone$ is a long expression. The $\dblb^{81}$ and $\rtwo^{16}$ factors suggest that the $q=3$ and $q=2$ loci respectively have multiplicities $81$ and $16$ within $\secy=3\secx^2+f\secz^4=0$. The $q=1$ multiplicity is therefore
\begin{equation}
\mone = 12\left(-K_B+[\secz]\right)\cdot\left(-K_B+[\secz]\right) - 81 \mthree - 16 \mtwo.
\end{equation}
This result agrees with the anomaly cancellation conditions in \eqref{eq:anom-cond}, as expected.

In most ways, the codimension two behavior parallels that for the \kmopr construction. However, the $q=3$ tuning derived here is slightly more general and admits matter spectra not possible in the \kmopr construction. For instance, consider a 6D F-theory model with base $\mathbb{P}^2$. The $q=3$ construction derived here admits a model in which $[\secz]=3H$, $[\dbla]=H$, and $[\dblb]=H$. The matter spectrum consists of a single $q=3$ hypermultiplet, 12 $q=2$ hypermultiplets, and $159$ $q=1$ hypermultiplets, a combination of charged matter that is not possible in the \kmopr construction. At the same time, there are seemingly consistent spectra that cannot be realized with the tuning presented here. As an example, for a $\mathbb{P}^2$ base and $[\secz] = 10H$, there is a SUGRA model with 4 $q=3$ hypermultiplets, 106 $q=2$ hypermultiplets and 8 $q=1$ hypermultiplets. But this spectrum cannot be realized with this $q=3$ tuning, as $\phibar$ would be ineffective. It would be interesting to determine whether there is an alternative $q=3$ construction realizing these missing matter spectra in future work. 

\subsection{UnHiggsings of the $q=3$ construction}
\label{subsec:g3unhiggsings}
Finally, let us summarize some of the potential ways that the $q=3$ construction can be unHiggsed to models with non-abelian groups. The general strategy is to consider ways to make the generating section ``vertical.'' Specifically, this entails making $\secz$ vanish. Since
\begin{align}
\secx &\equiv \cthree^2 \bmod{\secz} & \secy &\equiv \cthree^3 \bmod{\secz},
\end{align}
tuning $\secz\rightarrow 0$ makes the generating section equivalent to $[1:1:0]$, and the generating section coincides with the zero section. The different ways of unHiggsing described below correspond to different ways of tuning $\secz\rightarrow 0$. 
\paragraph{$\mathbf{U(1)\rightarrow SU(2)}$}
Field theoretically, giving a VEV to an adjoint of $\gsu(2)$ Higgses the $\gsu(2)$ symmetry down to $\gu(1)$. In many cases, F-theory $\gu(1)$ models exhibit the ``inverse'' of this Higgsing process, in which the $\gu(1)$ symmetry is enhanced to $\gsu(2)$. In the Morrison-Park model, taking $b\rightarrow 0$ often leads to a model with an $\gsu(2)$ tuned on $c_3$ \cite{morrison-park, mt-sections}. 
As noted in \cite{KleversTaylor}, the $\gu(1)$ symmetry in the \kmopr construction can also be enhanced to $\gsu(2)$ in many situations: taking the limit in which $\secz$ goes to zero (in a generic way) leads to an $\gsu(2)$ model with three-index symmetric ($\mathbf{4}$) matter. The $q=3$ tuning derived here admits a similar unHiggsing. We wish to make $\secz$ zero while keeping $\dbla$ and $\dblb$ generic. For a smooth base whose ring of sections can be treated as a UFD\footnote{Even though the divisor $\secz=0$ is singular, the base itself is taken to be smooth. Thus, the ring of sections on the base would be a UFD, but the quotient ring $\quotring{\secz}$ is not.}, the appropriate tunings are \cite{KleversTaylor} 
\begin{align}
\ba &\rightarrow 2\betaa\dblb & \bb &\rightarrow -\left(\betaa\dbla+\betac \dblb\right) & \bc &\rightarrow 2\betac\dbla.
\end{align}
This limit leads to a model equivalent to the $\gsu(2)$ model of \cite{kmrt} (up to simple redefinitions of the parameters), with the $\gsu(2)$ singularity tuned on
\begin{equation}
\cthree = \ta \dbla^3 + 3 \tb \dbla^2 \dblb + 3 \tc \dbla \dblb^2 + \td \dblb^3.
\end{equation}
$\cthree$ has triple point singularities, as expected: a $q=3$ model should enhance to an $\gsu(2)$ model with matter charged in the $\mathbf{4}$ representation, and $\mathbf{4}$ matter is supported at triple point singularities \cite{KleversTaylor}. It is reassuring that the $q=3$ tuning process motivated this cubic structure in $\cthree$ and reproduced the non-UFD structures encountered in the $\gsu(2)$ construction of \cite{kmrt}.

Note that $[\cthree]=-K_B + [\secz]$, which together with \eqref{eq:heightsec} implies that
\begin{equation}
h(\ratsec{s}) = 2[\cthree].
\end{equation}
This result reflects the known statement that $h(\ratsec{s})$ is equivalent to two times the homology class of the $\gsu(2)$ gauge divisor in the $\gu(1)\rightarrow \gsu(2)$ limit \cite{morrison-park,mt-sections}.

\paragraph{$\mathbf{U(1) \rightarrow SU(3)}$}
We also expect that, at least in certain situations, the $q=3$ construction can be enhanced to an $\gsu(3)$ model. In field theory, some $\gsu(N)$ models with appropriate charged matter spectra can be Higgsed in a particular fashion down to a $\gu(1)$ model with $q=N$ matter; if the $\gsu(N)$ is supported on a divisor with homology class $b_{SU(N)}$, the height $h(\ratsec{s})$ for the generating section should include a term of the form $N(N-1) b_{SU(N)}$ \cite{taylor-turner}. Some F-theory $\gu(1)$ models with $q=3$ matter should admit the corresponding unHiggsing process. The resulting $\gsu(3)$ tuning should be a standard UFD tuning when the $\gsu(3)$ charged hypermultiplets are in either the fundamental or adjoint representations.

For the $\gsu(3)$ unHiggsing, we still want to perform a tuning so that 
\begin{equation}
\secz = \ba \dbla^2 + 2 \bb \dbla \dbla + \bc \dblb^2
\end{equation} 
vanishes. In this case, we do not keep $\dblb$ generic, instead setting $\dblb$ and $\ba$ to $0$.\footnote{Alternatively, $\dbla$ and $\bc$ could be set to zero, leading to similar results.} The discriminant takes the form
\begin{equation}
\Delta =  \ta^2 \dbla^3 \Delta^\prime.
\end{equation}
Neither $f$ nor $g$ are proportional to $\ta$ after the tuning, so the resulting model has an $\gsu(2)$ symmetry tuned on $\ta$. Similarly, neither $f$ nor $g$ are proportional to $\dbla$, and since the split condition is satisfied, there is an $\gsu(3)$ symmetry tuned on $\dbla$. While there are codimension two (4,6) singularities at $\phibar=\dbla=0$, this issue can be avoided if we restrict our attention to situations in which $[\phibar]\cdot[\dbla]=0$. The tuning is a standard UFD tuning \cite{matter-singularities}, and there is no exotic matter in the spectrum, as expected.

We can compare $h(\ratsec{s})$ to the homology classes $[\ta]$ and $[\dbla]$.  Since $[\cthree] = [\ta]+3[\dbla]$,
\begin{equation}
h(\ratsec{s}) = 2[\ta] + 6[\dbla].
\end{equation}
The numerical factors of $2$ and $6$ agree with the $N(N-1)$ factor predicted by \cite{taylor-turner}.    

\section{Charge-4 models}
\label{sec:charge-4-models}

In this section, we derive and analyze an F-theory construction admitting $q=4$ matter. To the author's knowledge, this is the first published example of a $q=4$ F-theory model. In principle, such a model presumably could be derived using the normalized intrinsic ring, just as done for the $q=3$ case. However, given the algebraic complexity of the normalized intrinsic ring process, we use a somewhat indirect derivation. We deform a previous $\gu(1)\times \gu(1)$ construction admitting $(-2,-2)$ matter \cite{ckpt} and thereby Higgs the gauge group to a diagonal $\gu(1)$ with $q=4$ matter. The deformed construction has non-UFD structure tied to the presence of $q=4$ matter, which we examine after performing the deformation. However, we will not derive this structure from scratch. Note that this construction can likely be generalized and may not admit all of the possible F-theory $q=4$ spectra.\footnote{Evidence for this comes from unHiggsings of the $\gu(1)\times\gu(1)$ model in \cite{ckpt}. In particular, the $\gu(1)\times\gu(1)$ model can be enhanced to an $\gsu(3)$ model with symmetric matter, but there are $\gsu(3)$ constructions with symmetrics \cite{kmrt} that admit a wider variety of spectra. Since the enhanced $\gsu(3)$ model is not completely general, the $\gu(1)\times\gu(1)$ model in \cite{ckpt} (and the $q=4$ model after Higgsing) can likely be generalized in some way.} 

\subsection{Higgsing the U(1)$\times$U(1) construction}
\label{subsec:q4higgsing}

Our starting point is the $\gu(1)\times\gu(1)$ construction in \cite{ckpt}, which we refer to as the CKPT model. The discussion in \cite{ckpt} first describes this construction by embedding the elliptic curve in $\mathbb{P}^2$ with coordinates $[u:v:w]$:
\begin{multline}
p\equiv u\left(s_1 u^2 +s_2 u v+s_3 v^2 +s_5 u w+s_6 v w + s_8 w^2\right) \\
+ \left(a_1 v + b_1 w\right)\left(a_2 v + b_2 w\right)\left(a_3 v + b_3 w\right)=0\label{eq:p2-ecurve}.
\end{multline}
Here, the $s_i$, $a_i$, and $b_i$ are sections of line bundles on the base. There are three rational sections that are immediately obvious from \eqref{eq:p2-ecurve}:
\begin{align}
P &= \left[0:-b_1:a_1\right] & Q&=[0:-b_2: a_2] & R &= [0:-b_3: a_3]. 
\end{align}
Note that exchanges $a_2\leftrightarrow a_3$, $b_2\leftrightarrow b_3$ swap the sections $Q$ and $R$. One can then convert this construction to Weierstrass form; in \cite{ckpt}, $P$ is chosen to be the zero section, and the Mordell-Weil group is generated by $Q$ and $R$. The resulting $f$, $g$, and Weierstrass coordinates for $Q$ and $R$ are rather lengthy, so they are given in Appendix \ref{app:u1u1model} (with some minor corrections from \cite{ckpt}). The CKPT model supports $(-2,-2)$ matter at the $a_1=b_1=0$ loci. 

We now wish to Higgs the model and preserve a diagonal $\gu(1)$ so that the $(-2,-2)$ matter becomes $q=4$ matter after Higgsing. To implement this Higgsing at the F-theory level, we remove all instances of $a_2$, $a_3$, $b_2$, and $b_3$ through the following deformations:
\begin{align}
a_2 a_3 &\rightarrow d_0 & a_2 b_3 + a_3 b_2 &\rightarrow d_1 & b_2 b_3 &\rightarrow d_2. \label{eq:deformvars}
\end{align}
Arbitrary expressions involving $a_2$, $a_3$, $b_2$, and $b_3$ may not allow for this deformation, as the quantities being deformed are all invariant under $a_2\leftrightarrow a_3$, $b_2\leftrightarrow b_3$. But the description of the fibration in \eqref{eq:p2-ecurve} is consistent with \eqref{eq:deformvars}, taking the form 
\begin{multline}
p\equiv u\left(s_1 u^2 +s_2 u v+s_3 v^2 +s_5 u w+s_6 v w + s_8 w^2\right) \\
+ \left(a_1 v + b_1 w\right)\left(d_0 v^2 + d_1 v w + d_2 w^2\right)=0\label{eq:p2-ecurve-deformed}
\end{multline}
after the deformation. Note that this form is similar to the singular form of the \kmopr $q=3$ construction used in \cite{kmpopr,cvetic-z3}, but the zero section is not holomorphic. The $f$ and $g$ for the CKPT Weierstrass equation are also consistent with this deformation. The coordinates for $Q$ and $R$, either in $\mathbb{P}^2$ form or in Weierstrass form, cannot be deformed in this way; only expressions that are invariant under $a_2\leftrightarrow a_3$, $b_2\leftrightarrow b_3$ are compatible with \eqref{eq:deformvars}, and $a_2\leftrightarrow a_3$, $b_2\leftrightarrow b_3$ exchanges the sections $Q$ and $R$.\footnote{$Q\eplus(-R)$ is not compatible with \eqref{eq:deformvars} either: under $a_2\leftrightarrow a_3$, $b_2\leftrightarrow b_3$, the $[\secx:\secy:\secz]$ components transform to $[\secx:\secy:-\secz]$. One might wonder if the coordinates could be scaled by some expression that changes sign under $a_2\leftrightarrow a_3$, $b_2\leftrightarrow b_3$, so that the $\secz$ component would be invariant under the exchanges and could be deformed. However, this scaling would make $\secy$ change sign under the exchanges, and the components still could not be deformed.} However, $Q\eplus R$ is invariant under the exchange of $Q$ and $R$ and should be consistent with the deformation. This can be explicitly verified by calculating the coordinates of $Q\eplus R$ and then performing the deformation. Thus, $Q$ and $R$ are no longer rational sections after the deformation, but $Q\eplus R$ is a valid rational section. Instead of a rank-two Mordell-Weil group generated by $Q$ and $R$, we now have a rank-one Mordell-Weil group generated by $Q\eplus R$. We have therefore Higgsed the $\gu(1)\times\gu(1)$ gauge group down to a single diagonal $\gu(1)$. 

The Weierstrass equation and coordinates of the generating section after the deformation are lengthy. $f$, $g$, and the $\secz$ components are given in Appendix \ref{app:charge4appendix}, and the full model is given in the included Mathematica files discussed in Appendix \ref{app:mathematicafiles}. The homology classes of the parameters are summarized in Table \ref{tab:q4homology}. A particularly important part of the model is the $\secz$ component of the generating section, given by
\begin{multline}
\secz = \left(s_2 b_1 - s_5 a_1\right)\left(d_2 a_1^2-d_1 a_1 b_1 + d_0 b_1^2\right)^2-d_1\left(s_3 b_1^2 - s_6 a_1 b_1 +s_8 a_1^2\right)^2\\
+s_6\left(d_2 a_1^2-d_1 a_1 b_1 + d_0 b_1^2\right)\left(s_3 b_1^2 - s_6 a_1 b_1 +s_8 a_1^2\right)\\
+2 b_1\left(s_3 b_1^2 - s_6 a_1 b_1 +s_8 a_1^2\right)\left(b_1 d_1 s_3 - a_1 d_2 s_3 -b_1 d_0 s_6 +a_1 d_0 s_8\right).
\end{multline}
$\secz$ vanishes to order 4 at $a_1=b_1=0$, while the $\secx$ and $\secy$ components respectively vanish to orders 8 and 12. $a_1=b_1=0$ is also an $I_2$ locus. While the $I_2$ loci and their associated charged matter are discussed in \S\ref{subsec:q4matter}, we can immediately argue that the $a_1=b_1=0$ locus should support $q=4$ matter from the homomorphism properties of the Tate-Shioda map. The charge of matter at the $a_1=b_1=0$ locus is given by
\begin{equation}
q= \sigma(\ratsec{s})\cdot c,
\end{equation}
where $\sigma$ is the Tate-Shioda map, $\ratsec{s}$ is the generating section for the $q=4$ model, and $c$ is the extra fiber component at the $a_1=b_1=0$ locus. The Tate-Shioda map is a homomorphism, and (at least prior to the deformation) $\ratsec{s}= Q\eplus R$. Before the deformation, the charge with respect to $Q\eplus R$ is given by 
\begin{equation}
q= \sigma(Q\eplus R)\cdot c = \sigma(Q)\cdot c + \sigma(R)\cdot c.
\end{equation}
$\sigma(Q)\cdot c$ and $\sigma(R)\cdot c$ are simply the charges under the $\gu(1)\times\gu(1)$ group prior to the deformation, and the $a_1=b_1=0$ locus supports $(-2,-2)$ matter. The deformations preserve $Q\eplus R$ and affect neither $a_1$ nor $b_1$. Therefore, the charge after the deformation should be the sum of the $\gu(1)\times\gu(1)$ charged prior to Higgsing, implying that the $a_1=b_1=0$ locus supports $q=4$ matter.\footnote{The Higgsing argument suggests that, technically, $a_1=b_1=0$ matter should support $q=-4$ matter. However, the sign is unimportant, as the charged hypermultiplets in 6D consist of two half-hypermultiplets with opposite charges.}

\begin{table}
\begin{center}
\begin{tabular}{|c|c|}\hline
Parameter & Homology Class\\\hline
$a_1$ & $[a_1]$ \\
$b_1$ & $[b_1]$ \\
$d_0$ & $[\secz]+2 K_B-[a_1] - 3[b_1]$\\
$d_1$ & $[\secz]+2 K_B-2[a_1] - 2[b_1]$\\
$d_2$ & $[\secz]+2 K_B-3[a_1] - [b_1]$\\
$s_3$ & $-K_B + [a_1]-[b_1]$\\
$s_6$ & $-K_B$\\
$s_8$ & $-K_B - [a_1]+[b_1]$\\
$s_2$ & $-4K_B - [\secz]+2[a_1]+[b_1]$\\
$s_5$ & $-4K_B - [\secz]+[a_1]+2[b_1]$\\
$s_1$ & $-7K_B - 2[\secz]+3[a_1]+3[b_1]$\\\hline
\end{tabular}
\end{center}
\caption{Homology classes for parameters in the $q=4$ model. The classes are written in terms of the homology classes for $a_1$, $b_1$, and $\secz$ as well as the canonical class of the base $K_B$.}
\label{tab:q4homology}
\end{table}

\begin{figure}
\begin{center}
\includegraphics[scale=0.7]{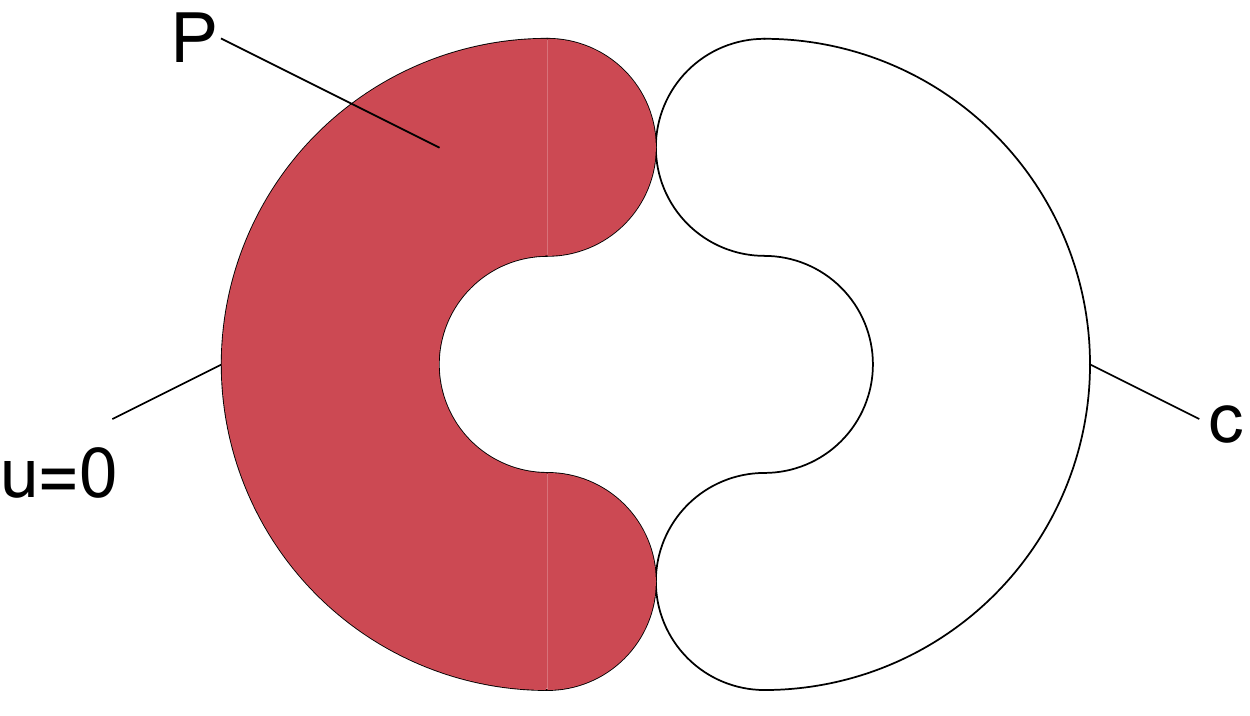}
\end{center}
\caption{Illustration of the fibers at $a_1=b_1=0$ loci for the $\mathbb{P}^2$ form of the fibrations. The fiber splits into two components. One component is given by $u=0$, and the other is denoted as $c$. The zero section $P$ wraps the $u=0$ component of the fiber, indicated by the filling of the $u=0$ component in the illustration.}
\label{fig:charge4res}
\end{figure}

The $\mathbb{P}^2$ form of the fibration provides an alternative way of seeing the presence of $q=4$ matter. We use a method based on the analysis in \cite{cvetic-z3}. In the $\mathbb{P}^2$ form, the tangent line at the zero section also hits the generating section $\ratsec{s}$. This tangent should be homologous to $u=0$, whose homology class we denote $U$. Therefore, the homology class of $\ratsec{s}$ should be $U-2\zerohomol$, where $\zerohomol$ is the homology class of the zero section. At $a_1=b_1=0$, \eqref{eq:p2-ecurve-deformed} factorizes into the form
\begin{equation}
u\left(s_1 u^2 +s_2 u v+s_3 v^2 +s_5 u w+s_6 v w + s_8 w^2\right)=0,
\end{equation}
indicating that the elliptic curve has split into two components. This situation is illustrated in Figure \ref{fig:charge4res}. Note that in the $\mathbb{P}^2$ form, the fibration is already smooth at $a_1=b_1=0$, although there are singularities at other codimension two loci in the base.\footnote{Ideally, we would use the fully resolved geometry to analyze the $a_1=b_1=0$ matter. But since the fibration is smooth at $a_1=b_1=0$, it suffices to consider the singular model in \eqref{eq:p2-ecurve-deformed}.} At $a_1=b_1=0$, the line $u=0$ becomes a full component, and it intersects the other component, which we denote as $c$, twice. The zero section $P$, meanwhile, becomes ill-defined at $a_1=b_1=0$ and wraps the $u=0$ component after being resolved; this behavior is identical to the behavior of the zero section in the CKPT construction \cite{ckpt}, as the deformations do not affect the zero section. $c$ is therefore the extra node, and the charge is given by
\begin{equation}
q = \sigma(\ratsec{s})\cdot c = \left(U - 3 \zerohomol\right)\cdot c.
\end{equation}
As just noted, the line $u=0$ intersects $c$ twice at $a_1=b_1=0$, and since the zero section wraps the same component, the zero section intersects twice as well. Therefore,
\begin{equation}
q = \left(U - 3 \zerohomol\right)\cdot c = 2- 3(2) = -4.
\end{equation}
Up to an unimportant negative sign, we see that the matter supported at $a_1=b_1=0$ has charge 4. 

\subsection{Structure of the charge-4 construction}

While we did not derive the $q=4$ construction using the normalized intrinsic ring, the expressions for $f$, $g$, and the components of the section hint at normalized intrinsic ring structure. Suppose we allow ourselves to freely divide by $a_1$, as would be the case if $a_1$ were a constant. Then, $\secz$ can be written in the suggestive form
\begin{equation}
\secz = \left(s_2 b_1 - s_5 a_1\right)\qthreea^2+\frac{1}{a_1}\left(s_6a_1 - 2 s_3 b_1\right)\qthreea\qthreeb-\frac{1}{a_1}\left(d_1 a_1 - 2 d_0 b_1\right)\qthreeb^2,
 \label{eq:zq4local}
\end{equation}
where
\begin{align}
\qthreea &= d_2 a_1^2 - d_1 a_1 b_1 + d_0 b_1^2 & \qthreeb &= s_8 a_1^2 - s_6 a_1 b_1 + s_3 b_1^2. \label{eq:q3loci} 
\end{align}
Like the $q=3$ $\secz$ component, the $q=4$ $\secz$ seems to admit a quadratic structure. However, the expressions $\qthreea$ and $\qthreeb$, which play the role of $\dbla$ and $\dblb$, are themselves quadratic in $a_1$ and $b_1$. From the discussion in \S\ref{subsec:q3matter}, the $\gu(1)$ symmetry in the $q=3$ construction can be unHiggsed to an $\gsu(3)$ symmetry tuned on either $\dbla$ or $\dblb$. At the same time, an $\gsu(3)$ model with matter charged in the symmetric representation ($\mathbf{6}$) can be Higgsed down to a $\gu(1)$ model with $q=4$ matter \cite{ckpt}. $\gsu(3)$ gauge groups supporting $\mathbf{6}$ matter are tuned on divisors with double point singularities \cite{sadov,matter-singularities}, so for the $q=4$ model, $\dbla$ and $\dblb$ should be replaced with some expressions having double point structure. This is exactly what is seen in \eqref{eq:zq4local}, as $\qthreea$ and $\qthreeb$ have the requisite quadratic structure. In fact, the height of the generating section is $6[\qthreeb] + 2([d_1]-[a_1]-[b_1])$, which displays the expected factor of 6 discussed in \S\ref{subsec:g3unhiggsings}.
The $((s_6a_1 - 2 s_3 b_1)/a_1$ and $(d_1 a_1 - 2 d_0 b_1)/a_1$ coefficients, meanwhile, are simply expressions for the normalized intrinsic ring parameters of $\qthreeb$ and $\qthreea$. \footnote{For example, compare these coefficients to \eqref{eq:normringdblb} and \eqref{eq:normringdbla}.} 

In fact, we can obtain the $f$ and $g$ for the $q=4$ Weierstrass model by starting with $f$ and $g$ for the $q=3$ model and making the replacements given in Table \ref{tab:q3q4conversion}. This observation provides further evidence that our construction supports $q=4$ matter, as the $\gu(1)\times\gu(1)$ Higgsings that give $q=4$ matter also lead to $q=3$ matter. If $a_1$ is constant (allowing us to divide freely by $a_1$), the highest charge supported by the model is $q=3$, and the two models should match. But the dictionary between the $q=3$ and $q=4$ constructions also suggests that the two Weierstrass models are birationally equivalent. In particular, $a_1^4 f$ and $a_1^6 g$ can be written in the form of a $q=3$ model \emph{without} division by $a_1$. Since $a_1^4 f\in -4K_B + 4[a_1]$ and $a_1^6 g\in -6K_B + 6[a_1]$, the Weierstrass model with $q=3$ structure is not a Calabi-Yau manifold unless $[a_1]$ is trivial. Thus, the $q=4$ model is birationally equivalent to the $q=3$ model, although the model in $q=3$ form does not satisfy the Calabi-Yau condition. This result seems to be a $q=4$ analogue of the statement in \cite{morrison-park-2016} that the Morrison-Park and the $q=3$ Weierstrass models are birationally equivalent. It is tempting to speculate that $\gu(1)$ models with $q>4$ should also be birationally equivalent to lower charge models; we leave a thorough investigation of this conjecture for future work.

\begin{table}
\begin{center}
\begin{tabular}{|c|c|}\hline
$q=3$ Parameter & Expression to obtain $q=4$ Model \\\hline
$\dbla$ & $\qthreea=d_2 a_1^2 - d_1 a_1 b_1 + d_0 b_1^2$\\
$\dblb$ & $\qthreeb=s_8 a_1^2 - s_6 a_1 b_1 + s_3 b_1^2$\\
$\ba$ & $b_1 s_2 - a_1 s_5$ \\
$\bb$ & $\frac{1}{2}(s_6 a_1 - 2 b_1 s_3)/a_1$\\
$\bc$ & $-(d_1 a_1 - 2 d_0 b_1)/a_1$\\
$\ta$ & $- s_1$\\
$\tb$ & $\frac{1}{3}s_2/a_1$\\
$\tc$ & $-\frac{1}{3} s_3 / a_1^2$\\
$\td$ & $d_0/a_1^2$\\
$\phibar$ & $12$ \\
$\hc$, $\hb$, $\ha$ & 0\\
$\lambdab$, $\lambdaa$ & 0\\
$f_2$ & 0\\
\hline
\end{tabular}
\end{center}
\caption{Replacements for converting the $q=3$ model to the $q=4$ model. Note that divisions by $a_1$ are required for the conversions.}
\label{tab:q3q4conversion}
\end{table}

In summary, the $q=3$ and $q=4$ models seem to be related, but the $q=4$ construction has some additional normalized intrinsic ring structure. It would be interesting to further examine the connections between the two constructions and use these patterns to obtain a more general $q=4$ form.

\subsection{Matter spectra}
\label{subsec:q4matter}
We now determine the codimension two $I_2$ singularities of the $q=4$ construction and the corresponding matter content. The results of this analysis are summarized in Tables \ref{tab:q4loci} and \ref{tab:q4matter}. There are two important aspects of the matter content analysis: the type of charge supported at an $I_2$ locus, and the multiplicity of matter fields with a particular charge. While the actual charge values are typically determined by resolving singularities, we instead use indirect methods to determine the charges, leaving a full resolution analysis for future work. However, we present more detailed calculations of the matter multiplicities. As in the $q=3$ matter analysis, we assume that we are working in six dimensions.

\begin{table}
\begin{center}
\begin{tabular}{|c|c|}\hline
Charge & $I_2$ Locus \\\hline
4 & $V(I_{q=4}) = \{a_1 = b_1 = 0\}$ \\
3 &$V(I_{q=3}) = \{\qthreea= \qthreeb = 0\}/V(I_{q=4})$ \\
2 & $V(I_{q=2}) =\{\cthreeqfour = \secz = 0\}/(V(I_{q=4})\cup V(I_{q=3}))$\\
1 & $V(I_{q=1}) =\{\secy = 3\secx^2 + f\secz^4 = 0\}/(V(I_{q=4})\cup V(I_{q=3})\cup V(I_{q=2}))$ \\\hline
\end{tabular}
\end{center}
\caption{$I_2$ loci for the $q=4$ construction along with the charges of the corresponding matter. Each locus is written as a variety $V$ associated to an ideal $I$ generated by two equations. $\qthreea$ and $\qthreeb$ are defined in \eqref{eq:q3loci}, while $\cthreeqfour$ is given in \eqref{eq:cthreeqfour}.}
\label{tab:q4loci}
\end{table}

\begin{table}
\begin{center}
\begin{tabular}{|c|c|}\hline
Charge &  Multiplicity \\\hline
4 &   $m_4= [a_1]\cdot[b_1]$\\
3 & $m_3 = ([\secz]+2K_B-[a_1]-[b_1])\cdot(-K_B + [a_1]+[b_1])-4[a_1]\cdot[b_1] $\\
2 &  $m_2 = [\secz]\cdot(-K_B+[\secz]) - 6 m_3 - 20 m_4$\\
1 &  $m_1 = 12(-K_B+[\secz])\cdot(-K_B+[\secz]) -16 m_2 - 81 m_3 - 256 m_4$\\\hline
\end{tabular}
\end{center}
\caption{Matter multiplicities for the $q=4$ construction.}
\label{tab:q4matter}
\end{table}

The codimension two $I_2$ loci are supported at the intersection of the divisors
\begin{align}
\secy &= 0 & 3\secx^2 + f \secz^4 = 0.
\end{align}
In principle, we could directly calculate the resultant of these two expressions and read off information about the matter spectrum. However, calculating this resultant is computationally complex, so we first consider the simpler problem of determining loci at which the section becomes ill-defined. Matter with $q\geq 2$ is supported at such loci, so this trick allows us to more quickly determine information about the matter content.

The important starting observation is that
\begin{align}
\secx &\equiv \cthreeqfour^2 \bmod \secz & \secy &\equiv \cthreeqfour^3 \bmod{\secz},
\end{align}
where\footnote{One can actually obtain this $\cthreeqfour$ expression by starting with the $\cthree$ of the $q=3$ construction, making the appropriate substitutions from Table \ref{tab:q3q4conversion}, and adding a term proportional to $\secz$ to remove all fractional terms. $\cthreeqfour$ vanishes to order 4 on $a_1=b_1=0$.}
\begin{multline}
\cthreeqfour = s_1 \qthreea^3 + \left(s_2 s_6 b_1 - s_2 s_8 a_1 - s_5 s_3 b_1 \right)\qthreea^2 +\left(2s_3 s_8 - s_6^2\right)\qthreea\qthreeb-\qthreeb^2\left(s_8 d_0 - s_6 d_1 + s_3 d_2\right)\\+a_1 s_6\left(d_0 s_8 b_1 - s_3 d_2 b_1 + s_6 d_2 a_1 - a_1 d_1 s_8\right)\qthreeb.\label{eq:cthreeqfour}
\end{multline}
The section is therefore ill-defined at $\cthreeqfour=\secz = 0$. This locus includes the loci $a_1=b_1=0$ and $\qthreea=\qthreeb=0$. By the homomorphism argument in \S\ref{subsec:q4higgsing}, the $a_1=b_1=0$ locus should support $q=4$ matter, with a $q=4$ multiplicity of $m_4=[a_1]\cdot[b_1]$. Meanwhile, we know that if $a_1$ is a constant, we recover a $q=3$ model with $q=3$ matter supported on the  $\qthreea=\qthreeb=0$ locus. This locus should still contribute $q=3$ matter even when $a_1$ is not a constant, as long as we exclude the $a_1=b_1=0$ locus. The $q=3$ locus is therefore $\{\qthreea=\qthreeb=0\}/\{a_1=b_1=0\}$. To count the $q=3$ multiplicities, we note that the resultant of $\qthreea$ and $\qthreeb$ with respect to $a_1$ takes the form
\begin{equation}
\Res_{a_1}(\qthreea,\qthreeb) = b_1^4 \rthree,
\end{equation}
with
\begin{equation}
r_3 = d_2^2 s_3^2 - d_1 d_2 s_3 s_6+d_0 d_2 s_6^2+d_1^2 s_3 s_8- 2 d_0 d_2 s_3 s_8 - d_0 d_1 s_6 s_8+d_0^2 s_8^2.
\end{equation}
The $b_1^4$ factor in the resultant suggests that $a_1=b_1=0$ is a degree 4 root of $\qthreea=\qthreeb=0$. In total, there are $[\qthreea]\cdot[\qthreeb]$ points in the $\qthreea=\qthreeb=0$ locus, so the $q=3$ multiplicity should be 
\begin{equation}
m_3=[\qthreea]\cdot[\qthreeb]-4[a_1]\cdot[b_1] = \left([\secz]+2K_B-[a_1]-[b_1]\right)\cdot\left(-K_B + [a_1]+[b_1]\right)-4[a_1]\cdot[b_1]. 
\end{equation}

Note that if we undo the deformations in \eqref{eq:deformvars}, $\qthreeb$ factorizes as $(a_1 b_2-a_2 b_1)(a_1 b_3 - a_3 b_1)$. This unHiggsing therefore splits the $\qthreea=\qthreeb=0$ locus into two loci: $(a_1 b_2-a_2 b_1) = \qthreea = 0$, and $(a_1 b_3 - a_3 b_1)=\qthreea=0$. In the original $\gu(1)\times\gu(1)$ model in \cite{ckpt}, these two loci support $(2,1)$ and $(1,2)$ matter, which are the types of charged matter that field theory considerations suggest should become $q=3$ matter after Higgsing. The match between these matter loci before and after Higgsing is further evidence that the $\qthreea=\qthreeb=0$ locus supports $q=3$ matter. 

The $q=2$ locus consists of the $\cthreeqfour = \secz =0$ points that do not support $q=4$ or $q=3$ matter. To calculate the $q=2$ multiplicity, we start with the $[\cthreeqfour]\cdot[\secz]$ intersection points and exclude those points corresponding to $q=4$ or $q=3$ matter. We therefore must examine the resultant of $\cthreeqfour$ and $\secz$ with respect to $a_1$, which is given by
\begin{equation}
\Res_{a_1}(\cthreeqfour,\secz) = b_1^{20} \rthree^6 d_2^3 \rtwo. 
\end{equation}
$r_2$ is a complicated, irreducible polynomial that we do not give here. The $b_1^{20}$ factor suggests that the $q=4$ locus is an degree 20 root of the system, while the $\rthree^6$ suggests that the $q=3$ locus is a degree 6 root of the system.\footnote{The $d_2^3$ factor is due to fact that the highest order $a_1$ terms in $\secz$ and $\cthreeqfour$ are both proportional to $d_2$. However, this does not correspond to a true locus at which $\secz$ and $\cthreeqfour$ both vanish.} Intriguingly, these numbers exactly match the $q^2(q^2-1)/12$ factors appearing in \eqref{eq:abeliangenus}. After removing the contributions from the $q=4$ and $q=3$ loci, we find that the $q=2$ multiplicity is given by
\begin{equation}
m_2 = [\secz]\cdot\left(-K_B + [\secz]\right) - 6 m_3 - 20 m_4.
\end{equation}
This result is in exact agreement with \eqref{eq:abeliangenus}, suggesting an F-theory interpretation of this anomaly equation. The $[\secz]\cdot\left(-K_B + [\secz]\right)$ reflects the fact that $q\geq 2$ matter is supported at places where the section components vanish; since
\begin{align}
\secx &\equiv \cthreeqfour^2 \bmod \secz & \secy &\equiv \cthreeqfour^3 \bmod{\secz},
\end{align}
the loci where the section components vanish are simply the $\cthreeqfour = \secz = 0$ loci. Meanwhile, the $q^2(q^2-1)/12$ factors represent the degree of the roots of the $\cthreeqfour = \secz = 0$ system.

$q=1$ matter is supported at the
\begin{equation}
\secy= 3\secx^2 + f \secz^4 =0 \label{eq:q4modelchargelocus}
\end{equation}
loci that do not support $q\geq2$ matter. $\secy$ and $3\secx^2+f \secz^4$ intersect at $12(-K_B+[\secz])^2$ points, but we must account for the the $q\geq 2$ loci before we can read off the $q=1$ multiplicity. We therefore need to calculate the multiplicities of the $q\geq 2$ loci within the locus described by \eqref{eq:q4modelchargelocus}. As in the $q=2$ and $q=3$ analyses, this information can be read off from the resultant with respect to $a_1$. In this case, calculating the resultant is computationally intensive if all parameters are allowed to be generic. We therefore evaluate the resultant for special cases in which some of the parameters are set to specific integer values. First, consider a situation where all parameters except $a_1$ and $b_1$ are set to specific integers. Then,
\begin{equation}
\Res_{a_1}(\secy,3\secx^2 + f \secz^4) \propto b_1^{256} \rtwo^{16},
\end{equation}
where $r_2$ is the same factor appearing in $\Res_{a_1}(\cthreeqfour,\secz)$ with the appropriate values for the parameters plugged in. This result suggests that the $a_1=b_1=0$ locus, which supports $q=4$ matter, has multiplicity $256$ within the \eqref{eq:q4modelchargelocus} locus, while the $q=2$ locus has multiplicity $16$. $\rthree$, which corresponds to the $q=3$ locus, does not depend on $b_1$, and when all parameters except $a_1$ and $b_1$ are set to integers, $\rthree$'s contribution to the resultant is simply an integer factor. To read off the $q=3$ multiplicity, we consider an alternative scenario in which all parameters except $a_1$ and $s_8$ are set to integers. Then,
\begin{equation}
\Res_{a_1}(\secy,3\secx^2 + f \secz^4) \propto \rthree^{81}\rtwo^{16},
\end{equation}
suggesting that the $q=3$ multiplicity is 81 and that the $q=2$ multiplicity is $16$. With these two results, we can now read off that
\begin{equation}
12(-K_B+[\secz])\cdot(-K_B+[\secz]) -16 m_2 - 81 m_3 - 256 m_4,
\end{equation}
exactly in agreement with the anomaly conditions in \eqref{eq:anom-cond}.

Finally, let us examine some possible ways of unHiggsing the $q=4$ construction. Of course, the $\gu(1)$ symmetry can be unHiggsed back to $\gu(1)\times\gu(1)$ by undoing the deformations in \eqref{eq:deformvars}. The model can then be further unHiggsed to an $\gsu(3)$ model supporting symmetric matter \cite{ckpt}. But there are other ways of unHiggsing the $\gu(1)$ symmetry to non-abelian gauge groups. As with the Morrison-Park and $q=3$ constructions, the general strategy is to tune parameters so that the generating section becomes vertical, coinciding with the zero section $[1:1:0]$. We therefore need to tune $\secz$ to vanish; $\secx$ and $\secy$ will then be a square and a cube of some expression, which can be scaled so that the generating section becomes $[1:1:0]$. 

In particular, let us restrict ourselves to unHiggsings in which $a_1$ is set to 0. Already, the discriminant is proportional to $b_1^2$, suggesting there is an $\gsu(2)$ tuned on $b_1=0$. The $\secz$ component (after rescaling the section coordinates by powers of $b_1$) takes the form
\begin{equation}
d_1 s_3^2 - s_6 d_0 s_3 + s_2 b_1 d_0^2,
\end{equation}
which is quadratic in $s_3$ and $d_0$. To make the section vertical, we should tune the above expression to zero. We cannot let both $d_0$ and $s_3$ be zero, as the discriminant then vanishes exactly. However, sending $s_3$ and $s_2$ to zero makes the discriminant proportional to $b_1^4 d_0^3 s_1^2$. $f$ and $g$ are not proportional to $b_1$, $d_0$, or $s_1$,implying that the enhanced model has an $\gsu(4)$ tuned on $b_1=0$, an $\gsu(3)$ tuned on $d_0=0$, and an $\gsu(2)$ tuned on $s_1=0$. The homology classes in Table \ref{tab:q4homology} imply that
\begin{equation}
h(\ratsec{s}) = 12[b_1]+6[d_0]+2[s_1].
\end{equation}
The coefficients for the homology classes supporting $\gsu(N)$ are given by $N(N-1)$, in agreement with the results from \S\ref{subsec:g3unhiggsings} and the expectations from \cite{taylor-turner}. 

An interesting question is whether the $q=4$ models admit unHiggsings to just an $\gsu(2)$ gauge group, like the Morrison-Park and $q=3$ constructions. This unHiggsing procedure would involve setting $\secz$ to be zero while keeping $a_1$ and $b_1$ generic. Presumably, the $\gsu(2)$ would be tuned on $\cthreeqfour=0$, which has a quadruple point singularity at $a_1=b_1=0$. So far, the author has not identified a way of actually performing this unHiggsing; in all cases considered, $\cthreeqfour$ factorizes, indicating the gauge group is product of non-abelian groups rather than a single $\gsu(2)$. However, a systematic investigation of all possible unHiggsings has not been performed. This issue has important implications for the F-theory swampland, which we discuss further in \S\ref{sec:conclusions}.


\section{Comments on $q>4$}

\label{sec:higher-charges}
We have seen that, in models with $q=3$ and $q=4$ matter, the components of the section vanish to higher orders at the loci supporting $q=3$ or $q=4$ matter. It is natural to speculate that similar behavior should occur for $q>4$ models. Without an explicit Weierstrass model, it is difficult to make definitive claims about $q>4$ matter. However, one can make conjectures about $q>4$ models by considering the behavior of the sections in a model admitting $q=1$ matter. Suppose that an F-theory model has a rank-one Mordell-Weil group with no additional non-abelian gauge groups. Let us denote the generating section as $\secgen$. If this F-theory model supports $q=1$ matter, there is some codimension two $I_2$ locus in which the elliptic curve splits into two components. One of these components, which we denote $c$, will not intersect the zero section, and because this locus supports $q=1$ matter,
\begin{equation}
\sigma(\secgen)\cdot c = 1.
\end{equation}
Using the elliptic curve addition law, we can construct sections $m\secgen$, where $m$ is some integer. From the homomorphism property of the Tate-Shioda map, the $m\secgen$ sections should satisfy
\begin{equation}
\sigma(m\secgen)\cdot c = m.
\end{equation}
The matter at this $I_2$ locus seems to have ``charge'' $m$ under the section $m\secgen$. Of course, $m\secgen$ does not generate the Mordell-Weil group for $|m|\neq 1$, and the matter supported at this locus does not truly have charge $m$. Nevertheless, the local behavior of $m\secgen$ likely mimics that of the generating section in a genuine $q=m$ model. We can therefore obtain some speculative insights into $q=m$ matter by examining the behavior of $m\secgen$.

This strategy was used in \cite{morrison-park} to anticipate the behavior of models supporting $q=2$ matter, and we use it here to conjecture about the behavior of sections admitting $q>2$ matter. We start with a simplified form of the Morrison-Park model that only supports $q=1$ matter \cite{morrison-park}. 
The Weierstrass model (in a chart where $z=1$) takes the form
\begin{equation}
\left(y+f_9\right)\left(y-f_9\right) = \left(x-f_6\right)\left(x^2 + f_6 x + \ftwelveh - 2 f_6^2 \right), \label{eq:simpmp-weierstrass}
\end{equation}
while the generating section is
\begin{equation}
\secgen: [x:y:z]=[f_6:f_9:1]. \label{eq:simpmp-section}
\end{equation}
There are $I_2$ singularities at $f_9=\ftwelveh=0$ that, according to the analysis in \cite{morrison-park}, support $q=1$ matter. Our goal here is to use the elliptic curve addition law to calculate the $m\secgen$ sections and examine their behavior at $f_9=\ftwelveh=0$. For example, the $2\secgen$ section takes the form
\begin{align}
2\secgen: [x:y:z] = [\ftwelveh^2 - 8 f_6 f_9^2: -\ftwelveh^3+12 f_6 \ftwelveh f_9^2 - 8 f_9^4: 2 f_9].\label{eq:twicesec}
\end{align}
The ($\secz$, $\secx$, $\secy$) components vanish to orders $(1,2,3)$ at $f_9=\ftwelveh=0$, in agreement with the known behavior of sections at $q=2$ loci.

\begin{table}
\begin{center}
\begin{tabular}{|c|c|c|c|}\hline
$m$ & $\secz$ Order of Vanishing & $\secx$ Order of Vanishing & $\secy$ Order of Vanishing\\ \hline
$2$ & $1$ & $2$ & $3$ \\
$3$ & $2$ & $4$ & $7$ \\
$4$ & $4$ & $8$ & $12$ \\
$5$ & $6$ & $12$ & $19$ \\
$6$ & $9$ & $18$ & $27$ \\
$7$ & $12$ & $24$ & $37$ \\
$8$ & $16$ & $32$ & $48$ \\
$9$ & $20$ & $40$ & $61$ \\
$10$ & $25$ & $50$ & $75$ \\
$11$ & $30$ & $60$ & $91$ \\\hline
\end{tabular}
\end{center}
\caption{Orders of vanishing for the $m\secgen$ components at $f_9=\ftwelveh=0$. These are calculated using the simplified Morrison-Park model described by Equations \eqref{eq:simpmp-weierstrass} and \eqref{eq:simpmp-section}. The orders of vanishing for $m=2$, $3$, and $4$ agree with the behavior of the generating sections in models admitting $q=2$, $3$, and $4$ matter. One can therefore conjecture that the components of sections in models supporting $q=m$ matter would vanish to the orders listed here at the $q=m$ loci.}
\label{tab:ordvanishmsec}
\end{table}

For $m=3$, we again see behavior in line with the known $q=3$ models. The components of $3\secgen$, given by
\begin{align}
\secz =& \ftwelveh^2-12 f_6 f_9^2\\
\secx=& \ftwelveh^4 f_6+8 \ftwelveh^3 f_9^2-24 \ftwelveh^2 f_6^2 f_9^2-96 \ftwelveh f_6 f_9^4+144 f_6^3 f_9^4+64 f_9^6\\
\secy=& 3 \ftwelveh^6 f_9-60 \ftwelveh^4 f_6 f_9^3+96 \ftwelveh^3 f_9^5+144 \ftwelveh^2 f_6^2 f_9^5-1152 \ftwelveh f_6 f_9^7\notag\\ &+64 \left(27 f_6^3 f_9^7+8 f_9^9\right),
\end{align}
vanish to orders $(2,4,7)$ at $f_9=\ftwelveh=0$, just like the components of the generating section for the $q=3$ construction in \S\ref{sec:charge-3-models}. The $(\secz,\secx,\secy)$ components for $4\secgen$ vanish to orders $(4,8,12)$. The generating section for the $q=4$ construction vanishes to these same orders at the $q=4$ loci, giving further credence to the idea that this construction truly supports $q=4$ matter. 

Table \ref{tab:ordvanishmsec} summarizes the orders of vanishing for the $m\ratsec{s}$ sections at $f_9=\ftwelveh=0$. As expected, the $m>2$ section components show singular behavior, with the $\secz$, $\secx$,and $\secy$ components vanishing to orders greater than $1$. Given that the behavior of the $m=2,3,4$ sections agrees with the behavior of the known $q=2,3,4$ models, one can conjecture that the generating section components for $q>4$ models will also vanish to these orders at the $q=m$ loci. In fact, the cases presented in Table \ref{tab:ordvanishmsec} suggest patterns in the orders of vanishing. For even $m$, the orders of vanishing for $(\secz, \secx, \secy)$ seem to be given by
\begin{equation}
\left(\frac{m^2}{4}, \frac{2 m^2}{4}, \frac{3m^2}{4} \right).
\end{equation}
Meanwhile, the $(\secz, \secx, \secy)$ orders of vanishing for odd values of $m$ seem to be given by
\begin{align}
\left(\frac{m^2-1}{4},\frac{2(m^2-1)}{4},\frac{3(m^2-1)}{4}+1\right)
\end{align}
These patterns have been verified for the $m\secgen$ sections with $m\leq 26$. 

It would be interesting to investigate whether the patterns hold for all values of $m$, both in the simplified Morrison-Park form and in actual $q=m$ models. Perhaps the expressions could be proven with a better understanding of the resolutions at the $f_9=\ftwelveh=0$ loci. These questions are left for future work. But if these orders of vanishing are correct, this information may be useful for inferring features of the $q=m$ Weierstrass models. Recall that in \S\ref{sec:charge-3-models}, the $q=3$ Weierstrass model could be derived with the knowledge that $\secz$ vanishes to order 2 on the $q=3$ loci. In the same way, one might hope that the orders of vanishing determine the Weierstrass model's structure in a predictable fashion, allowing for a systematic derivation of $q>4$ models. These patterns could also give a quick way of detecting the presence of $q>2$ matter. Regardless of the type of charge supported, charged matter in a $\gu(1)$ model (that is not also charged under some additional non-abelian symmetry) occurs at an $I_2$ locus, so examining the discriminant does not provide an immediate way of reading off the charge. But if the behavior of the $\secz$ component can distinguish between the different charges, one may be able to at least guess the charge content of a model without the need for an explicit resolution.

\section{Conclusions and future directions}

To summarize, we have constructed $\gu(1)$ F-theory models admitting both $q=3$ and $q=4$ matter. In both cases, all of the section components vanish to orders higher than 1 at the $q=3, 4$ matter loci. As a result, the Weierstrass models have non-UFD structure that deviates from the standard Morrison-Park form. With the aid of the normalized intrinsic ring, we were able to find the appropriate non-UFD structures for the $q=3$ matter and systematically derive a generalization of the $q=3$ construction described in \cite{kmpopr}. A class of $q=4$ constructions were also constructed, although the models were found by deforming the earlier $\gu(1)\times\gu(1)$ construction in \cite{ckpt}. Nevertheless, the $q=4$ construction shows signs of normalized intrinsic ring structure as well. We finally discussed some conjectures regarding models with $q>4$ matter.

A natural direction for future work is to search for models admitting $q>4$ matter. There are a few different strategies that may give new insights into this issue. Just as deforming a $\gu(1)\times\gu(1)$ model led to $q=4$ matter, deforming models with multiple $\gu(1)$ factors could lead to larger charges. This process would likely require an initial model with somewhat exotic matter charged under multiple $\gu(1)$ factors. For instance, the possible Higgsings of the $\gu(1)^3$ construction in \cite{cvetic-u1cubed} cannot give $q>4$ matter, although they can produce $q=3$ and $q=4$ matter. Alternatively, one could obtain large charges by Higgsing models with non-abelian symmetry.  \cite{taylor-turner} gives examples of the field-theoretic Higgsing processes that could produce $q>4$ matter. However, it can be difficult to identify the deformations of F-theory models corresponding to a specific Higgsing. A better understanding of the F-theory realizations of Higgsing processes, particularly Higgsing on adjoints, would be helpful to develop concrete methods for $q>4$ models. There is the possibility of building $q>4$ models from scratch, although the algebraic complexity of the models discussed here suggests this approach may be unwieldy. Based on the $q=3$ derivation in \S\ref{sec:charge-3-models}, we would likely need some knowledge of the $q>4$ singularity structures. Analyses similar to \S5.1 of \cite{morrison-park} or \S\ref{sec:higher-charges} here could provide the necessary insights to construct $q>4$ Weierstrass models. At the very least, such efforts could illustrate the local behavior of sections at loci supporting arbitrary charges.

But there are interesting questions about $q=3$ and $q=4$ models as well. On the one hand, neither of the Weierstrass models discussed here admit the full range of matter spectra consistent with the anomaly equations in \eqref{eq:anom-cond}, suggesting that there may be generalizations of these constructions. In particular, the $q=4$ construction can almost certainly be extended in some way. The models should also be subjected to a more thorough resolution analysis. Resolutions of the $q=3$ construction should be similar to resolutions of the \kmopr construction in \cite{kmpopr}, and the analysis of the $q=4$ matter loci in \S\ref{subsec:q4higgsing} paints a rough picture of the behavior of the section there. Nevertheless, a more rigorous analysis of the codimension two singularities would be helpful for confirming the matter analysis presented here. It would also be useful to count the uncharged hypermultiplets in these models, possibly with the techniques used in \cite{yinan-u1}. 

Meanwhile, the $q=3$ and $q=4$ sections discussed here (as well as the $q>4$ sections in \S\ref{sec:higher-charges}) have components that exhibit singular behavior, raising the question of whether the sections themselves are singular. Preliminary indications suggest that the sections are indeed singular. One can describe the section using a system of equations: in addition to equations describing the elliptic fibration, the system would include equations such as $x \secz^2-\secx z^2=0$, where $\secx$ and $\secz$ refer to the section components. One can then use the Jacobian condition to determine loci where the section is singular. Of course, the elliptic fibration needs to be resolved at the relevant codimension two singularities, and the section needs to be resolved to account for loci where the section components vanish. After the resolution procedure, the section may wrap a component of an $I_2$ fiber, as described previously. An initial analysis indicates that, at the $q\geq3$ loci, many of the sections described here are singular at the intersections between the $I_2$ fiber components. This information would be important for comparing the models presented here to the results in \cite{f-theory-rational}. However, a more thorough analysis should be performed to understand any possible singularities in these sections. It would be interesting to explore these issues in future work, possibly in the context of a broader analysis of singular sections.

The $q=3$ and $q=4$ models also offer avenues to explore F-theory physics. When a model has $q>2$ matter, the anomaly cancellation conditions \eqref{eq:anom-cond} do not uniquely determine the spectrum, even if one fixes $h(\ratsec{s})$ and $K_B$. In non-abelian contexts where this situation occurs, there are matter transitions connecting the vacua with different matter spectra \cite{matter-in-transition}. Abelian F-theory models should also exhibit such transitions, which would change the charge content of the theory without changing the gauge group or other parts of the spectrum. Because the $\gsu(2)$ construction in \cite{kmrt} admits matter transitions, the $q=3$ construction here, which can be unHiggsed to this same $\gsu(2)$ model, should admit matter transitions as well. Seeing transitions involving $q=4$ matter would probably require some generalization of the construction given here. Because abelian symmetries manifest themselves differently than non-abelian symmetries in F-theory, $\gu(1)$ transitions would likely give a new understanding of these models. Matter transitions could also be used to derive $q>4$ models. For instance, an $\gsu(4)$ gauge group with $\mathbf{10}$ matter (and a suitable number of adjoints) can be Higgsed down to a $\gu(1)$ with $q=6$ matter through a process similar to the $\gsu(4)\rightarrow \gu(1)$ Higgsing discussed in \S\ref{subsec:q4matter}. $\gsu(4)$ models have matter transitions that change the amount of $\mathbf{10}$ matter, implying that the corresponding $\gu(1)$ models should have transitions that change the amount of $q=6$ matter. In particular, one can start with an $\gsu(4)$ model without $\mathbf{10}$ matter and use the transitions to grow $\mathbf{10}$ matter \cite{kmrt}. Thus, the explicit $\gsu(4)$ transition could potentially be used to reverse engineer a $\gu(1)$ matter transition that generates a $q=6$ model from a known $q=4$ model. 

Unhiggsing $q=4$ models could also be an important check of the swampland statement in \cite{kmrt} that certain non-abelian representations, including the $\mathbf{5}$ representation of $\gsu(2)$, cannot be realized in F-theory. Field theoretically, if an $\gsu(2)$ symmetry is Higgsed down to $\gu(1)$, the presence of $\mathbf{5}$ matter would lead to $q=4$ matter after Higgsing. An examination of unHiggsings of the $q=4$ construction is therefore important, as an enhancement to an $\gsu(2)$ model with $\mathbf{5}$ would invalidate the statement. However, it is crucial to note that the existence of a $q=4$ F-theory model does not by itself guarantee the existence of an $\gsu(2)$ model with $\mathbf{5}$ matter. For instance, the would-be $\gsu(2)$ divisor may factor into multiple components, much like the situation observed in \cite{Baume-Cvetic-Lawrie-Lin}. Alternatively, the resulting $\gsu(2)$ Weierstrass model may have codimension two (4,6) singularities \cite{kmrt}. So far, the author has not identified a way of achieving this $\gsu(2)$ enhancement, but a complete investigation of all possible unHiggsings has not been done.  

Finally, the investigations here hint at a deeper interpretation of the section components that should be understood better. The $\secz$ component seems to be the defining feature of the $q=3$ construction, and the anomaly equation \eqref{eq:abeliangenus} seems to manifest itself through the section components. Understanding the physical meaning of the section components may provide new insights into abelian F-theory models. For instance, \cite{Braun-Collinucci-Valandro,MayorgaPena-Valandro} analyze $\gu(1)$ models, including the Morrison-Park construction and the original $q=3$ construction in \cite{kmpopr}, in the Sen limit. Similar Type IIB investigations could elucidate the role played by the section components. In any case, a more physical description of the models discussed here may inform efforts to find $\gu(1)$ models admitting larger charges.

\label{sec:conclusions}

\acknowledgments I would especially like to thank Washington Taylor, not only for his numerous comments on the technical aspects of this work, but also for his incredible support and encouragement while this work was being completed. Additionally, I would like to thank Mirjam Cvetic, Denis Klevers, Craig Lawrie, Dave Morrison, Paul Oehlmann, and Andrew Turner for helpful discussions. I am also grateful to Samuel Monnier, Gregory Moore, and Daniel Park for sharing work prior to publication. This work is supported by the Office of High Energy Physics of U.S. Department of Energy under Contract Number DE-SC0012567.

\appendix

\section{Mathematica files}
\label{app:mathematicafiles}
There are two Mathematica files, \texttt{Charge3Model.nb} and \texttt{Charge4Model.nb}, that respectively contain expressions for the $q=3$ and $q=4$ constructions. These files are contained as ancillary files in the arXiv submission and can be obtained by downloading the gzipped source of the submission. Each file contains the $f$ and $g$ of the Weierstrass model (assigned to the variables \texttt{f} and \texttt{g}) and the $\secx$, $\secy$, and $\secz$ components of the generating section (assigned to the variables \texttt{x}, \texttt{y}, and \texttt{z}). Because some of the parameters have typographical features, such as subscripts or macrons, that are not easily used in Mathematica, the Mathematica variable names may be slightly different than the parameter names used here. Tables \ref{tab:charge3mathematica} and \ref{tab:charge4mathematica} give the dictionaries between the model parameters and Mathematica variables.

\begin{table}
\begin{center}
\begin{tabular}{|c|c|}\hline
Parameter & Mathematica Variable\\\hline
$\dbla$ & $\eta$\texttt{a}\\
$\dblb$ & $\eta$\texttt{b}\\
$\bc$ & \texttt{b0}\\
$\bb$ & \texttt{b1}\\
$\ba$ & \texttt{b2}\\
$\phibar$ & $\varphi$ \\
$\td$ & \texttt{t0}\\
$\tc$ & \texttt{t1}\\
$\tb$ & \texttt{t2}\\
$\ta$ & \texttt{t3}\\
$\hc$ & \texttt{h0}\\
$\hb$ & \texttt{h1}\\
$\ha$ & \texttt{h2}\\
$\lambdab$ & $\lambda$\texttt{0}\\
$\lambdaa$ & $\lambda$\texttt{1}\\
$f_2$ & \texttt{f2}\\\hline\end{tabular}
\end{center}
\caption{Mathematica variables corresponding to the parameters in the $q=3$ model.}
\label{tab:charge3mathematica}
\end{table}

\begin{table}
\begin{center}
\begin{tabular}{|c|c|}\hline
Parameter & Mathematica Variable\\\hline
$a_1$ & \texttt{a1}\\
$b_1$ & \texttt{b1}\\
$d_0$ & \texttt{d0}\\
$d_1$ & \texttt{d1}\\
$d_2$ & \texttt{d2}\\
$s_1$ & \texttt{s1}\\
$s_2$ & \texttt{s2}\\
$s_3$ & \texttt{s3}\\
$s_5$ & \texttt{s5}\\
$s_6$ & \texttt{s6}\\
$s_8$ & \texttt{s8}\\\hline 
\end{tabular}
\end{center}
\caption{Mathematica variables corresponding to the parameters in the $q=4$ model.}
\label{tab:charge4mathematica}
\end{table}

\section{Charge-3 expressions}
\label{app:charge3appendix}
The components of the section (in Weierstrass form) are given by
\begin{align}
\secz =& \ba \dbla^2 + 2 \bb \dbla\dblb + \bc \dblb^2, \\
\secx =& \cthree^2 -\frac{1}{432}\left[72\left(\ha\dbla^2+2\hb\dbla\dblb+\hc\dblb^2\right)-\phibar^2\left(\bb^2-\bc\ba\right)\right]\secz^2\notag\\
&+ \frac{\phibar}{6}\Big[\left(\ba\dbla+\bb\dblb\right)\left(\tb\dbla^2+2\tc\dbla\dblb+\td\dblb^2\right)\notag\\
&\phantom{+ \frac{\phibar}{6}\qquad}-\left(\bb\dbla+\bc\dblb\right)\left(\ta\dbla^2+2\tb\dbla\dblb+\tc\dblb^2\right)\Big]\secz,\\
\secy =& \cthree^3-\frac{1}{72}\cthree\left[18\left(\ha\dbla^2+2\hb\dbla\dblb+\hc\dblb^2\right)-\phibar^2\left(\bb^2-\bc\ba\right)\right]\secz^2+\frac{1}{2}\left(\lambdaa\dbla+\lambdab\dblb\right)\secz^4\notag\\
&+ \frac{\phibar}{4}\cthree\Big[\left(\ba\dbla+\bb\dblb\right)\left(\tb\dbla^2+2\tc\dbla\dblb+\td\dblb^2\right)\notag\\
&\phantom{+ \frac{\phibar}{6}\qquad}-\left(\bb\dbla+\bc\dblb\right)\left(\ta\dbla^2+2\tb\dbla\dblb+\tc\dblb^2\right)\Big]\secz,\notag\\
&+ \frac{\phibar}{48}\left[\left(\ha\dbla+\hb\dblb\right)\left(\bb\dbla+\bc\dblb\right)-\left(\hb\dbla+\hc\dblb\right)\left(\ba\dbla+\bb\dblb\right)\right]\secz^3\notag\\
&+ \frac{\phibar^2}{96}\left[\ba\left(\tc\dbla+\td\dblb\right)-2\bb\left(\tb\dbla+\tc\dblb\right)+\bc\left(\ta\dbla+\tb\dblb\right)\right]\secz^3,
\end{align}
where
\begin{equation}
\cthree = \ta\dbla^3+3\tb\dbla^2\dblb+3\tc\dbla\dblb^2+\td\dblb^3.
\end{equation}

For the Weierstrass model, $f$ is given by
\begin{multline}
f = -\frac{1}{48}\left[\ha\dbla^2+2\hb\dbla\dblb+\hc\dblb^2+\frac{\phibar^2}{36}\left(\bb^2-\bc\ba\right)\right]^2 \\
- \frac{\phibar}{24}\Bigg[\left(\hb\dbla+\hc\dblb\right)\left(\ta\dbla^2+2\tb\dbla\dblb+\tc\dblb^2\right)\\
\shoveright{-\left(\ha\dbla+\hb\dblb\right)\left(\tb\dbla^2+2\tc\dbla\dblb+\td\dblb^2\right)\Bigg]} \\
- \frac{\phibar^2}{48}\left[\left(\tb\dbla+\tc\dblb\right)^2-\left(\tc\dbla+\td\dblb\right)\left(\ta\dbla+\tb\dblb\right)\right]\\
+ \frac{1}{1728}\phibar^3\Bigg[\left(\ba \td - 2 \bb \tc+\bc \tb\right)\left(\ba\dbla+\bb\dblb\right)\\
\shoveright{-\left(\ba \tc-2\bb\tb+\bc\ta\right)\left(\bb \dbla+\bc\dblb\right)\Bigg]}\\
+\left(\lambdaa\dbla+\lambdab\dblb\right)\cthree 
- \frac{1}{576}\phibar^2\left(\ha\bc-2\hb\bb+\hc\ba\right)\secz \\
+ \frac{1}{12}\phibar\left[\lambdab\left(\ba \dbla + \bb \dblb\right)-\lambdaa\left(\bb \dbla+\bc \dblb\right)\right]\secz+ f_2\secz^2.
\end{multline}
Meanwhile, $g$ is given by
\begin{equation}
g = g_0 + g_1 \secz + g_2 \secz^2,
\end{equation}
with
\begin{multline}
g_0 = \frac{1}{864}\hcombo^3 +\frac{3}{864}\phibar\hcombo\left[\left(\hb\dbla+\hc\dblb\right)\left(\taua \dbla + \taub \dblb\right)-\left(\ha\dbla+\hb\dblb\right)\left(\taub \dbla + \tauc \dblb\right)\right] \\
+ \frac{3}{864}\phibar^2\hcombo\tausq + \frac{1}{864}\phibar^3\taucu+\frac{\phibar^2}{576}\left(\ha\tauc-2\hb\taub+\hc\taua\right)\cthree - \frac{1}{12}\hcombo\left(\lambdaa\dbla+\lambdab \dblb\right)\cthree\\
-\frac{1}{12}\phibar\left[\lambdab\left(\taua \dbla+\taub\dblb\right)-\lambdaa\left(\taub\dbla+\tauc\dblb\right)\right]\cthree-f_2\cthree^2\\
+\left(\frac{\phibar^2}{144}\right)^2\left[\frac{1}{4}\left(\ba\tauc-2\bb\taub+\bc\taua\right)^2-\left(\bb^2-\bc\ba\right)\tausq\right]\\
-\frac{\phibar^3}{(144)^2}\hcombo\Bigg[\left(\ba \td - 2 \bb \tc+\bc \tb\right)\left(\ba\dbla+\bb\dblb\right)\\
-\left(\ba \tc-2\bb\tb+\bc\ta\right)\left(\bb \dbla+\bc\dblb\right)\Bigg],
\end{multline}
\begin{multline}
g_1 = -\frac{\phibar}{6}\left[\left(\taub\dbla+\tauc\dblb\right) \left(\ba \dbla + \bb \dblb\right) - \left(\taua\dbla+\taub\dblb\right)\left(\bb \dbla + \bc \dblb\right)\right]f_2 \\
+\frac{1}{3}\left(\frac{\phibar}{48}\right)^2\hcombo\left(\ba\hc-2\bb\hb+\bc\ha\right)\\
+\frac{1}{288}\left(\frac{\phibar^2}{6}\right)^2\left[\ba\left(\tc^2-\td\tb\right)-\bb\left(\tc\tb-\td\ta\right)+\bc\left(\tb^2-\tc\ta\right)\right]\\
 - \frac{1}{144}\phibar \hcombo\left[\lambdab\left(\ba\dbla+\bb\dblb\right)-\lambdaa\left(\bb\dbla+\bc\dblb\right)\right]\\
+\frac{3}{288}\phibar^2\left(\lambdaa\dbla+\lambdab\dblb\right)\left(\ba\tauc-2\bb\taub+\bc\taua\right)\\
\shoveleft{+\frac{1}{72}\phibar^2\Bigg[\lambdab\left(\bb\left(\taua\dbla+\taub\dblb\right)-\ba\left(\taub\dbla+\tauc\dblb\right)\right)}\\
\shoveright{-\lambdaa\left(\bc\left(\taua\dbla+\taub\dblb\right)-\bb\left(\taub\dbla+\tauc\dblb\right)\right)\Bigg]}\\
\shoveleft{+\frac{\phibar^3}{3456}\Bigg[\left(\ha\td-2\hb\tc+\hc\tb\right)\left(\ba\dbla+\bb\dblb\right)}\\
\shoveright{-\left(\ha\tc-2\hb\tb+\hc\ta\right)\left(\bb\dbla+\bc\dblb\right)\Bigg]}\\
\shoveleft{-\frac{\phibar^3}{6912}\Bigg[\left(\ba\td-2\bb\tc+\bc\tb\right)\left(\ha\dbla+\hb\dblb\right)}\\
-\left(\ba\tc-2\bb\tb+\bc\ta\right)\left(\hb\dbla+\hc\dblb\right)\Bigg],
\end{multline}
and
\begin{multline}
g_2 = \frac{1}{4}\left(\lambdaa\dbla+\lambdab\dblb\right)^2 + \frac{1}{6}f_2\left[\hcombo -\frac{\phibar^2}{24}\left(\bb^2-\bc\ba\right)\right]+\frac{\phibar^2}{2304}\left(\hb^2-\hc\ha\right)\\
+\frac{1}{48}\phibar\Bigg[\lambdab\left(\ha\dbla + \hb \dblb\right)-\lambdaa\left(\hb\dbla+\hc\dblb\right)\Bigg].
\end{multline}
In the $g_i$ expressions, we have used
\begin{align}
\hcombo &= \ha \dbla^2 + 2 \hb\dbla\dblb+\hc\dblb^2 + \frac{\phibar^2}{36}\left(\bb^2-\bc\ba\right),\\
\taua &= \ta\dbla+\tb\dblb,\\
\taub &= \tb\dbla+\tc\dblb,\\
\tauc &= \tc\dbla+\td\dblb,\\
\tausq &= \taub^2-\taua\tauc,
\end{align}
and
\begin{multline}
\taucu = \frac{1}{2}\Bigg[-\left(2\tb^3-3 \tc\tb\ta+\td\ta^2\right)\dbla^3-3\left(\tc \tb^2-2\tc^2\ta+\td\tb\ta\right)\dbla^2\dblb\\
+3\left(\tc^2\tb - 2 \td\tb^2+\td \tc \ta\right)\dbla \dblb^2+\left(2\tc^3-3 \td \tc \tb + \td^2 \ta\right)\dblb^3\Bigg].
\end{multline}

\section{Charge-4 expressions}
\label{app:charge4appendix}
\subsection{$\mathbb{P}^2$ form}
In the $\mathbb{P}^2$ form of the elliptic fibration, in which the elliptic fiber is described via an embedding in $\mathbb{P}^2$, the $q=4$ model is
\begin{multline}
p\equiv u\left(s_1 u^2 +s_2 u v+s_3 v^2 +s_5 u w+s_6 v w + s_8 w^2\right) \\
+ \left(a_1 v + b_1 w\right)\left(d_0 v^2 + d_1 v w + d_2 w^2\right)=0,
\end{multline}
where $[u:v:w]$ are the $\mathbb{P}^2$ coordinates. The zero section has components
\begin{equation}
[u:v:w] = [0:-b_1:a_1],
\end{equation}
while the generating section has components $[u:v:w]$ with 
\begin{align}
u =& \qthreea\Big[\left(s_2 b_1 - s_5 a_1\right)\qthreea^2\\
&\phantom{\qthreea\Big[}-\qthreeb\left(d_0 s_6 b_1^2+2d_2 s_3 a_1 b_1 -s_3 d_1 b_1^2-d_2 s_6 a_1^2 - 2 d_0 s_8 a_1 b_1+s_8 d_1 a_1^2\right)\Big],\notag\\
v=& -s_1 b_1 \qthreea^3 + s_5 \qthreea^2 \qthreeb + d_2\left(s_3 b_1 - s_6 a_1\right)\qthreeb^2 - s_8\left(d_0 b_1 - d_1 a_1\right)\qthreeb^2 ,\notag\\
w=& s_1 a_1 \qthreea^3 -s_2 \qthreea^2 \qthreeb - s_3\left(d_1 b_1 - d_2 a_1\right)\qthreeb^2+d_0\left(s_6 b_1 - s_8 a_1\right)\qthreeb^2.
\end{align}
$\qthreea$ and $\qthreeb$ are defined as
\begin{align}
\qthreea &= d_2 a_1^2 - d_1 a_1 b_1 + d_0 b_1^2 & \qthreeb &= s_8 a_1^2 - s_6 a_1 b_1 + s_3 b_1^2.\label{eq:appq3loci} 
\end{align}

\subsection{Weierstrass form}
The $f$ and $g$ in Weierstrass form are given by
\begin{multline}
f=-\frac{1}{3} \left(s_5^2-3 s_1 s_8\right) \left(a_1^2 \left(d_1^2-3 d_0 d_2\right)-a_1 b_1 d_0 d_1+b_1^2 d_0^2\right)\\
-\frac{1}{3} \left(s_2^2-3 s_1 s_3\right) \left(a_1^2 d_2^2+b_1^2 \left(d_1^2-2 d_0 d_2\right)\right)\\
+\frac{1}{6} (2 s_2 s_5-3 s_1 s_6) \left(a_1^2 d_1 d_2+a_1 b_1 \left(d_1^2-2 d_0 d_2\right)+b_1^2 d_0 d_1\right)\\
+\frac{1}{6} (a_1 d_1+b_1 d_0) \left(2 b_1 d_2 \left(s_2^2-3 s_1 s_3\right)-3 s_2 s_6 s_8+s_5 \left(2 s_3 s_8+s_6^2\right)\right)\\
+a_1 d_0 \left(b_1 d_2 (3 s_1 s_6-2 s_2 s_5)+s_2 s_8^2-\frac{s_5 s_6 s_8}{2}\right)\\
+\frac{1}{6} (a_1 d_2+b_1 d_1) \left(s_3 (2 s_2 s_8-3 s_5 s_6)+s_2 s_6^2\right)\\
+\frac{1}{2} b_1 d_2 s_3 (2 s_3 s_5-s_2 s_6)-\frac{1}{48} \left(s_6^2-4 s_3 s_8\right)^2
\end{multline}
\filbreak
\begin{multline}
g=\frac{1}{864} \left(s_6^2-4 s_3 s_8\right)^3-\frac{1}{2} \left(d_0 d_2^3 a_1^4+b_1^3 d_0 \left(d_1^3-3 d_0 d_1 d_2\right) a_1\right) s_1^2\\
+\frac{1}{4} \left(d_2^2 \left(d_1^2-2 d_0 d_2\right) a_1^4+b_1^2 \left(d_1^4-6 d_0^2 d_2^2-4 d_0 d_2 \left(d_1^2-2 d_0 d_2\right)\right) a_1^2+b_1^4 d_0^2 \left(d_1^2-2 d_0 d_2\right)\right) s_1^2\\
+\frac{1}{27} \left(\left(d_1^3-3 d_0 d_1 d_2\right) b_1^3+a_1^3 d_2^3\right) s_2 \left(9 s_1 s_3-2 s_2^2\right)\\
+\frac{1}{18} \left(d_1 d_2^2 a_1^3+b_1^2 \left(d_1^3-3 d_0 d_1 d_2\right) a_1+b_1^3 d_0 \left(d_1^2-2 d_0 d_2\right)\right) \left(\left(2 s_2^2-3 s_1 s_3\right) s_5-3 s_1 s_2 s_6\right)\\
+\frac{1}{18} \left(d_0 d_1 d_2 a_1^3+b_1 d_0 \left(d_1^2-2 d_0 d_2\right) a_1^2+b_1^2 d_0^2 d_1 a_1\right) \left(2 s_5^3-9 s_1 s_8 s_5+9 b_1 d_2 s_1^2\right)\\
\shoveleft{+\frac{1}{72} \left(d_1 d_2 a_1^2+b_1 \left(d_1^2-2 d_0 d_2\right) a_1+b_1^2 d_0 d_1\right)}\\ \times\Big[4 b_1 d_2 s_2 \left(2 s_2^2-9 s_1 s_3\right)+s_6 \left(s_6 (2 s_2 s_5+3 s_1 s_6)-12 s_3 s_5^2\right)\\
\shoveright{+4 \left(s_2 s_3 s_5-3 \left(s_2^2-5 s_1 s_3\right) s_6\right) s_8\Big]}\\
+\frac{1}{18} \left(d_2 \left(d_1^2-2 d_0 d_2\right) a_1^3+b_1 \left(d_1^3-3 d_0 d_1 d_2\right) a_1^2+b_1^3 d_0^2 d_1\right) \left(s_2 \left(2 s_5^2-3 s_1 s_8\right)-3 s_1 s_5 s_6\right)\\
+\frac{2}{9} \left(d_0 d_2^2 a_1^3+b_1^2 d_0 \left(d_1^2-2 d_0 d_2\right) a_1\right) \left(3 s_1 s_5 s_6+s_2 \left(3 s_1 s_8-2 s_5^2\right)\right)\\
+a_1^2 d_0^2 \left(-\frac{3}{2} b_1^2 d_2^2 s_1^2+\frac{1}{4} s_8^2 \left(s_5^2-4 s_1 s_8\right)+\frac{2}{9} b_1 d_2 s_5 \left(9 s_1 s_8-2 s_5^2\right)\right)\\
+\frac{1}{36} \left(\left(d_1^2-2 d_0 d_2\right) b_1^2+a_1^2 d_2^2\right) \left(3 \left(3 s_5^2-8 s_1 s_8\right) s_3^2+\left(4 s_2^2 s_8-3 s_6 (2 s_2 s_5+s_1 s_6)\right) s_3+2 s_2^2 s_6^2\right)\\
+\frac{1}{24} b_1 d_2 s_3 \left(6 b_1 d_2 s_3 \left(s_2^2-4 s_1 s_3\right)+(s_2 s_6-2 s_3 s_5) \left(s_6^2-4 s_3 s_8\right)\right)\\
\shoveleft{+\frac{1}{36} \left(d_0 d_2 a_1^2+b_1 d_0 d_1 a_1\right) \Big[\left(s_6^2+2 s_3 s_8\right) s_5^2+18 s_2 s_6 s_8 s_5-6 \left(s_2^2+2 s_1 s_3\right) s_8^2}\\
\shoveright{+4 b_1 d_2 \left(\left(2 s_2^2-3 s_1 s_3\right) s_5-3 s_1 s_2 s_6\right)-33 s_1 s_6^2 s_8\Big]}\\
-\frac{1}{54} \left(\left(d_1^3-3 d_0 d_1 d_2\right) a_1^3+b_1^3 d_0^3\right) \left(4 s_5^3+9 s_1 (3 b_1 d_2 s_1-2 s_5 s_8)\right)\\
\shoveleft{+\frac{1}{72} a_1 d_0 \Big[16 b_1^2 s_2 \left(9 s_1 s_3-2 s_2^2\right) d_2^2}\\
+6 b_1 \left(s_6 \left(6 s_3 s_5^2+s_6 (9 s_1 s_6-8 s_2 s_5)\right)+2 \left(3 \left(s_2^2+2 s_1 s_3\right) s_6-8 s_2 s_3 s_5\right) s_8\right) d_2\\
\shoveright{+3 s_8 (s_5 s_6-2 s_2 s_8) \left(s_6^2-4 s_3 s_8\right)\Big]}\\
+\frac{1}{18} \left(d_0 d_1 a_1^2+b_1 d_0^2 a_1\right) \left(2 b_1 d_2 \left(s_2 \left(2 s_5^2-3 s_1 s_8\right)-3 s_1 s_5 s_6\right)-3 s_8 \left(s_6 s_5^2+(s_2 s_5-6 s_1 s_6) s_8\right)\right)\\
-\frac{1}{72} (b_1 d_1+a_1 d_2) \left(12 b_1 d_2 s_3 \left(s_2 s_3 s_5+\left(s_2^2-6 s_1 s_3\right) s_6\right)+\left(s_6^2-4 s_3 s_8\right) \left(s_2 s_6^2+s_3 (2 s_2 s_8-3 s_5 s_6)\right)\right)\\
\shoveleft{+\frac{1}{72} (b_1 d_0+a_1 d_1) \Big[2 b_1 d_2 \left(-6 \left(s_5^2+2 s_1 s_8\right) s_3^2+\left(2 s_8 s_2^2+18 s_5 s_6 s_2-33 s_1 s_6^2\right) s_3+s_2^2 s_6^2\right)}\\
\shoveright{-\left(s_6^2-4 s_3 s_8\right) \left(s_5 \left(s_6^2+2 s_3 s_8\right)-3 s_2 s_6 s_8\right)\Big]}\\
\shoveleft{+\frac{1}{36} \left(\left(d_1^2-2 d_0 d_2\right) a_1^2+b_1^2 d_0^2\right) \Big[2 \left(s_6^2+2 s_3 s_8\right) s_5^2-6 s_2 s_6 s_8 s_5}\\
+8 b_1 d_2 \left(-2 s_5 s_2^2+3 s_1 s_6 s_2+3 s_1 s_3 s_5\right)-3 s_8 \left(s_1 \left(s_6^2+8 s_3 s_8\right)-3 s_2^2 s_8\right)\Big].
\end{multline}

The $\secz$ component of the generating section is
\begin{multline}
\secz = \left(s_2 b_1 - a_1 s_5\right)\qthreea^2\\
-\qthreeb\left(d_0 s_6 b_1^2+2d_2 s_3 a_1 b_1 -s_3 d_1 b_1^2-d_2 s_6 a_1^2 - 2 d_0 s_8 a_1 b_1+s_8 d_1 a_1^2\right),
\end{multline}
with $\qthreea$ and $\qthreeb$ defined as in \eqref{eq:appq3loci}. The $\secx$ and $\secy$ components are lengthy and are not given here. However, they are included in the Mathematica notebooks described in Appendix \ref{app:mathematicafiles}. 

\section{U(1)$\mathbf{\times}$U(1) expressions}
\label{app:u1u1model}
The below formulas are for the $\gu(1)\times\gu(1)$ model of \cite{ckpt}, with some minor typos corrected. For the Weierstrass model, the $f$ and $g$ are given by
\begin{multline}
f= -\frac{1}{48} \left(s_6^2-4 s_3 s_8\right)^2+\frac{1}{2} b_1 b_2 b_3 s_3 (2 s_3 s_5-s_2 s_6)-\frac{1}{3} \left(a_3^2 b_1^2 b_2^2+a_2^2 b_1^2 b_3^2+a_1^2 b_2^2 b_3^2\right) \left(s_2^2-3 s_1 s_3\right)\\
+\frac{1}{6} (a_1 b_2 b_3+a_2 b_1 b_3+a_3 b_1 b_2) \left(s_2 s_6^2+s_3 (2 s_2 s_8-3 s_5 s_6)\right)\\
+\frac{1}{6} (b_1 a_2 a_3+b_2 a_1 a_3+b_3 a_1 a_2) \left(2 b_1 b_2 b_3 \left(s_2^2-3 s_1 s_3\right)-3 s_2 s_6 s_8+s_5 \left(s_6^2+2 s_3 s_8\right)\right)\\
+\frac{1}{6} \left(a_2 a_3^2 b_1^2 b_2+a_2^2 a_3 b_1^2 b_3+a_1 a_3^2 b_1 b_2^2+a_1^2 a_3 b_2^2 b_3+a_1 a_2^2 b_1 b_3^2+a_1^2 a_2 b_2 b_3^2\right) (2 s_2 s_5-3 s_1 s_6)\\
-\frac{1}{3} \left(a_2^2 a_3^2 b_1^2+a_1^2 a_2^2 b_3^2+b_2^2 a_1^2 a_3^2-a_1 a_2 a_3^2 b_1 b_2-a_1 a_2^2 a_3 b_1 b_3-a_1^2 a_2 a_3 b_2 b_3\right) \left(s_5^2-3 s_1 s_8\right)\\
+a_1 a_2 a_3 \left(b_1 b_2 b_3 (3 s_1 s_6-2 s_2 s_5)+s_2 s_8^2-\frac{s_5 s_6 s_8}{2}\right),
\end{multline}
\filbreak
\begin{multline}
g=\frac{1}{864} \left(s_6^2-4 s_3 s_8\right)^3+\frac{1}{24} b_1 b_2 b_3 s_3 \left((s_2 s_6-2 s_3 s_5) \left(s_6^2-4 s_3 s_8\right)+6 b_1 b_2 b_3 s_3 \left(s_2^2-4 s_1 s_3\right)\right)\\
\shoveleft{-\frac{1}{72} (b_2 b_3 a_1+b_3 b_1 a_2+b_1 b_2 a_3)}\\
\shoveright{\times\left(12 b_1 b_2 b_3 s_3 \left(s_2 s_3 s_5+\left(s_2^2-6 s_1 s_3\right) s_6\right)+\left(s_6^2-4 s_3 s_8\right) \left(s_2 s_6^2+s_3 (2 s_2 s_8-3 s_5 s_6)\right)\right)}\\
+\frac{1}{36} \left(b_1^2 b_2^2 a_3^2+b_1^2 b_3^2 a_2^2+b_2^2 b_3^2 a_1^2\right) \left(3 \left(3 s_5^2-8 s_1 s_8\right) s_3^2+\left(4 s_2^2 s_8-3 s_6 (2 s_2 s_5+s_1 s_6)\right) s_3+2 s_2^2 s_6^2\right)\\
+\frac{1}{72} (b_1 a_2 a_3+b_2 a_1 a_3+b_3 a_1 a_2) \Big[2 b_1 b_2 b_3 \left(-6 \left(s_5^2+2 s_1 s_8\right) s_3^2+\left(2 s_8 s_2^2+18 s_5 s_6 s_2-33 s_1 s_6^2\right) s_3+s_2^2 s_6^2\right)\\
\shoveright{-\left(s_6^2-4 s_3 s_8\right) \left(s_5 \left(s_6^2+2 s_3 s_8\right)-3 s_2 s_6 s_8\right)\Big]}\\
+\frac{1}{27} \left(b_1^3 b_2^3 a_3^3+b_1^3 b_3^3 a_2^3+b_2^3 b_3^3 a_1^3\right) s_2 \left(9 s_1 s_3-2 s_2^2\right)\\
\shoveleft{+\frac{1}{72} \left(b_1^2 b_2 a_2 a_3^2+b_1^2 b_3 a_2^2 a_3+a_1 a_2^2 b_1 b_3^2+a_1^2 a_2 b_2 b_3^2+a_1^2 a_3 b_2^2 b_3+a_1 a_3^2 b_1 b_2^2\right) }\\
\times\Big[4 b_1 b_2 b_3 s_2 \left(2 s_2^2-9 s_1 s_3\right)+s_6 \left(s_6 (2 s_2 s_5+3 s_1 s_6)-12 s_3 s_5^2\right)\\
\shoveright{+4 \left(s_2 s_3 s_5-3 \left(s_2^2-5 s_1 s_3\right) s_6\right) s_8\Big]}\\
\shoveleft{+\frac{1}{72} a_1 a_2 a_3 \Big[16 b_1^2 b_2^2 b_3^2 s_2 \left(9 s_1 s_3-2 s_2^2\right)}\\
+6 b_1 b_2 b_3 \left(s_6 \left(6 s_3 s_5^2+s_6 (9 s_1 s_6-8 s_2 s_5)\right)+2 \left(3 \left(s_2^2+2 s_1 s_3\right) s_6-8 s_2 s_3 s_5\right) s_8\right)\\
\shoveright{+3 s_8 (s_5 s_6-2 s_2 s_8) \left(s_6^2-4 s_3 s_8\right)\Big]}\\
+\frac{1}{18} \left(a_2^3 a_3 b_1^3 b_3^2+a_3^3 a_2 b_1^3 b_2^2+a_3^3 a_1 b_1^2 b_2^3+a_1^3 a_3 b_2^3 b_3^2+a_1^3 a_2 b_2^2 b_3^3+a_2^3 a_1 b_1^2 b_3^3\right) \left(\left(2 s_2^2-3 s_1 s_3\right) s_5-3 s_1 s_2 s_6\right)\\
\shoveleft{+\frac{1}{36} \left(b_1^2 a_2^2 a_3^2+b_2^2 a_1^2 a_3^2+b_3^2 a_1^2 a_2^2\right) \Big[8 b_1 b_2 b_3 \left(-2 s_5 s_2^2+3 s_1 s_6 s_2+3 s_1 s_3 s_5\right)}\\
\shoveright{+2 \left(s_6^2+2 s_3 s_8\right) s_5^2-6 s_2 s_6 s_8 s_5-3 s_8 \left(s_1 \left(s_6^2+8 s_3 s_8\right)-3 s_2^2 s_8\right)\Big]}\\
\shoveleft{+\frac{1}{36} \left(a_1^2 a_2 a_3 b_2 b_3+a_2^2 a_1 a_3 b_1 b_3+a_3^2 a_1 a_2 b_1 b_2\right) }\\
\times\Big[4 b_1 b_2 b_3 \left(\left(2 s_2^2-3 s_1 s_3\right) s_5-3 s_1 s_2 s_6\right)+\left(s_6^2+2 s_3 s_8\right) s_5^2\\
\shoveright{+18 s_2 s_6 s_8 s_5-6 \left(s_2^2+2 s_1 s_3\right) s_8^2-33 s_1 s_6^2 s_8\Big]}\\
+\frac{1}{18} \left(b_2 b_3^3 a_1^3 a_2^2+b_1 b_3^3 a_2^3 a_1^2+b_2^3 b_3 a_1^3 a_3^2+b_1 b_2^3 a_3^3 a_1^2+b_1^3 b_3 a_2^3 a_3^2+b_1^3 b_2 a_3^3 a_2^2\right) \left(s_2 \left(2 s_5^2-3 s_1 s_8\right)-3 s_1 s_5 s_6\right)\\
+\frac{2}{9} \left(b_1^2 b_2^2 a_1 a_2 a_3^3+b_1^2 b_3^2 a_1 a_3 a_2^3+b_2^2 b_3^2 a_2 a_3 a_1^3\right) \left(3 s_1 s_5 s_6+s_2 \left(3 s_1 s_8-2 s_5^2\right)\right)\\
\shoveleft{+\frac{1}{18} \left(b_1 a_1 a_2^2 a_3^2+b_2 a_2 a_1^2 a_3^2+b_3 a_3 a_1^2 a_2^2\right)}\\
\shoveright{\times\left(2 b_1 b_2 b_3 \left(s_2 \left(2 s_5^2-3 s_1 s_8\right)-3 s_1 s_5 s_6\right)-3 s_8 \left(s_6 s_5^2+(s_2 s_5-6 s_1 s_6) s_8\right)\right)}\\
+\frac{1}{4} \left(b_1^4 b_2^2 a_2^2 a_3^4+b_1^4 b_3^2 a_2^4 a_3^2+b_2^2 b_3^4 a_1^4 a_2^2+b_1^2 b_3^4 a_2^4 a_1^2+b_2^4 b_3^2 a_1^4 a_3^2+b_1^2 b_2^4 a_3^4 a_1^2\right) s_1^2\\
-\frac{1}{2} \left(b_1^3 b_2^3 a_1 a_2 a_3^4+b_2^3 b_3^3 a_2 a_3 a_1^4+b_1^3 b_3^3 a_1 a_3 a_2^4\right) s_1^2\\
-\frac{1}{54} \left(a_1^3 a_2^3 b_3^3+a_2^3 a_3^3 b_1^3+a_1^3 a_3^3 b_2^3\right) \left(9 s_1 (3 b_1 b_2 b_3 s_1-2 s_5 s_8)+4 s_5^3\right)\\
\shoveleft{+\frac{1}{18} \left(b_1^2 b_2 a_1 a_2^2 a_3^3+b_1 b_2^2 a_2 a_1^2 a_3^3+b_1^2 b_3 a_1 a_3^2 a_2^3+b_3^2 b_1 a_1^2 a_3 a_2^3+b_2^2 b_3 a_2 a_3^2 a_1^3+b_2 b_3^2 a_3 a_2^2 a_1^3\right)}\\
\shoveright{\times\left(9 b_1 b_2 b_3 s_1^2+2 s_5^3-9 s_1 s_8 s_5\right)}\\
+(a_1 a_2 a_3)^2 \left(-\frac{3}{2} b_1^2 b_2^2 b_3^2 s_1^2+\frac{2}{9} b_1 b_2 b_3 s_5 \left(9 s_1 s_8-2 s_5^2\right)+\frac{1}{4} s_8^2 \left(s_5^2-4 s_1 s_8\right)\right).
\end{multline}

There are two generating sections, $Q$ and $R$. The Weierstrass components of $Q$ are
\begin{align*}
z_Q &= a_1 b_2 - a_2 b_1,\stepcounter{equation}\tag{\theequation}\\
x_Q &= b_1^2 b_2^2 s_3^2-b_1 b_2 (a_2 b_1+a_1 b_2) s_3 s_6+\frac{1}{12} \left(a_2^2 b_1^2+a_1^2 b_2^2\right) \left(s_6^2+8 s_3 s_8\right)+\frac{1}{6} a_1 a_2 b_1 b_2 \left(5 s_6^2+4 s_3 s_8\right)\\
&+\frac{1}{3} (a_2 b_1-a_1 b_2)^2 ((2 a_3 b_1 b_2-a_2 b_1 b_3-a_1 b_2 b_3) s_2+(-a_2 a_3 b_1-a_1 a_3 b_2+2 a_1 a_2 b_3) s_5)\\
&-a_1 a_2 (a_2 b_1+a_1 b_2) s_6 s_8+a_1^2 a_2^2 s_8^2,\stepcounter{equation}\tag{\theequation}\\
y_Q &=-b_1^3 b_2^3 s_3^3+\frac{1}{2} (a_2 b_1-a_1 b_2)^4 (a_3 b_1-a_1 b_3) (-a_3 b_2+a_2 b_3) s_1-\frac{1}{2} b_2^4 b_3 s_3 s_5 a_1^4-\frac{1}{2} a_3 b_2^4 s_2 s_8 a_1^4\\
&+\frac{1}{2} a_2 b_2^3 b_3 (s_5 s_6+s_2 s_8) a_1^4+\frac{1}{2} a_2 a_3 b_2^3 s_5 s_8 a_1^4-a_2^2 b_2^2 b_3 s_5 s_8 a_1^4+\frac{1}{2} a_3 b_1 b_2^4 (s_3 s_5+s_2 s_6) a_1^3\\
&+a_2 a_3 b_1 b_2^3 (s_2 s_8-s_5 s_6) a_1^3+\frac{1}{2} b_2^3 s_3 (b_1 b_2 b_3 s_2+s_6 s_8) a_1^3-\frac{1}{2} a_2^2 a_3 b_1 b_2^2 s_5 s_8 a_1^3\\
&+a_2^3 \left(-s_8^3+2 b_1 b_2 b_3 s_5 s_8\right) a_1^3-\frac{1}{2} a_2^2 b_2 \left(b_1 b_2 b_3 (s_5 s_6+s_2 s_8)-3 s_6 s_8^2\right) a_1^3\\
&+\frac{1}{2} a_2 b_2^2 \left(2 b_1 b_2 b_3 (s_3 s_5-s_2 s_6)-s_8 \left(s_6^2+2 s_3 s_8\right)\right) a_1^3-a_3 b_1^2 b_2^4 s_2 s_3 a_1^2-\frac{1}{2} a_2 a_3 b_1^2 b_2^3 (s_3 s_5+s_2 s_6) a_1^2\\
&-\frac{1}{2} b_1 b_2^3 s_3 \left(s_6^2+2 s_3 s_8\right) a_1^2+\frac{1}{2} a_2 b_1 b_2^2 \left(s_6^3+5 s_3 s_8 s_6-b_1 b_2 b_3 s_2 s_3\right) a_1^2+a_2^2 a_3 b_1^2 b_2^2 (2 s_5 s_6-s_2 s_8) a_1^2\\
&-\frac{1}{2} a_2^3 a_3 b_1^2 b_2 s_5 s_8 a_1^2-a_2^4 b_1^2 b_3 s_5 s_8 a_1^2-\frac{1}{2} a_2^3 b_1 \left(b_1 b_2 b_3 (s_5 s_6+s_2 s_8)-3 s_6 s_8^2\right) a_1^2\\
&-a_2^2 b_1 b_2 \left(b_1 b_2 b_3 (s_3 s_5-2 s_2 s_6)+s_8 \left(2 s_6^2+s_3 s_8\right)\right) a_1^2+2 a_2 a_3 b_1^3 b_2^3 s_2 s_3 a_1+\frac{3}{2} b_1^2 b_2^3 s_3^2 s_6 a_1\\
&-\frac{1}{2} a_2^2 a_3 b_1^3 b_2^2 (s_3 s_5+s_2 s_6) a_1+a_2^3 a_3 b_1^3 b_2 (s_2 s_8-s_5 s_6) a_1+\frac{1}{2} a_2^4 b_1^3 b_3 (s_5 s_6+s_2 s_8) a_1\\
&-a_2 b_1^2 b_2^2 s_3 \left(2 s_6^2+s_3 s_8\right) a_1+\frac{1}{2} a_2^4 a_3 b_1^3 s_5 s_8 a_1+\frac{1}{2} a_2^2 b_1^2 b_2 \left(s_6^3+5 s_3 s_6 s_8-b_1 b_2 b_3 s_2 s_3\right) a_1\\
&+\frac{1}{2} a_2^3 b_1^2 \left(2 b_1 b_2 b_3 (s_3 s_5-s_2 s_6)-s_8 \left(s_6^2+2 s_3 s_8\right)\right) a_1-a_2^2 a_3 b_1^4 b_2^2 s_2 s_3\\
&-\frac{1}{2} a_2^4 b_1^4 b_3 s_3 s_5+\frac{3}{2} a_2 b_1^3 b_2^2 s_3^2 s_6+\frac{1}{2} a_2^3 a_3 b_1^4 b_2 (s_3 s_5+s_2 s_6)-\frac{1}{2} a_2^4 a_3 b_1^4 s_2 s_8-\frac{1}{2} a_2^2 b_1^3 b_2 s_3 \left(s_6^2+2 s_3 s_8\right)\\
&+\frac{1}{2} a_2^3 b_1^3 s_3 (b_1 b_2 b_3 s_2+s_6 s_8)\stepcounter{equation}\tag{\theequation}
\end{align*}
The components of the $R$ section can be found by taking the $Q$ components and performing $a_2\leftrightarrow a_3$, $b_2 \leftrightarrow b_3$.


\bibliographystyle{JHEP}
\bibliography{references}






\end{document}